\documentclass[]{interact}

\usepackage{graphicx}
\usepackage{epstopdf}
\usepackage[caption=false]{subfig}

\usepackage[numbers,sort&compress,merge]{natbib}
\bibpunct[, ]{[}{]}{,}{n}{,}{,}
\renewcommand\bibfont{\fontsize{10}{12}\selectfont}

\theoremstyle{plain}

\theoremstyle{definition}

\theoremstyle{remark}

\usepackage{xcolor}

\begin{document}


\title{Anomalous behaviour of transport properties in a supercooled water-glycerol mixture}

\author{
\name{Alberto Zaragoza,\textsuperscript{a,b,e} Rajat Kumar,\textsuperscript{a} Jose Martin Roca, \textsuperscript{b} Prabir Khatua,\textsuperscript{c} Valeria Molinero,\textsuperscript{a} Frédéric Caupin \textsuperscript{d} and Chantal Valeriani\textsuperscript{b} \thanks{CONTACT C. Valeriani. Email: cvaleriani@ucm.es}
}
\affil{\textsuperscript{a}Department of Chemistry The University of Utah, Salt Lake City, UT 84112-0850; \textsuperscript{b}Departamento de Estructura de la Materia, F\'isica T\'ermica y Electr\'onica, Universidad Complutense de Madrid, 28040 Madrid, Spain; 
\textsuperscript{c} Department of Chemistry,GITAM School of Science Bengaluru, Karnataka, IN 561203 India.
\textsuperscript{d} Institute Lumière matière, Université Claude Bernard Lyon 1, CNRS, Institut Universitaire de France, F-69622 Villeurbanne, France.
\textsuperscript{e} Departamento de Matemáticas y Ciencias de Datos, Universidad San Pablo-CEU, CEU Universities, Madrid 28003, Spain}
}

\maketitle

\begin{abstract}
Glycerol acts as a natural cryoprotectant by depressing the temperature of ice nucleation and slowing down the dynamics of water mixtures. In this work we characterize dynamics -diffusion, viscosity, and hydrogen-bond dynamics- as well as density anomaly and structure of water mixtures with 1\% to 50\% w/w glycerol at low temperatures via molecular dynamics simulations using all-atom and coarse-grained models. Simulations reveal distinct violations of the Stokes-Einsten relation in the low temperature regime for water and glycerol. Deviations are positive for water at all concentrations, and positive for glycerol in very dilute solutions but turning negative in concentrated ones. The all-atom and coarse-grained models reveal an unexpected crossover in the dynamics of the 1\% and 10 \% w/w glycerol at the lowest simulated temperatures. This crossover manifests in the diffusion coefficients of water and glycerol, as well as in the viscosity and  lifetime of hydrogen-bonds in water. We interpret that the crossover originates on the opposing dependence with glycerol concentration of the two factors controlling the solutions' slow-down: the increase in tetrahedrally coordinated
water and the dynamics and clustering of the glycerol molecules. We anticipate that this dynamic crossover will also occur for solution of water with other polyols. 

\end{abstract}


\section{Introduction}

Water-polyol mixtures have unique properties and wide-ranging applications. These solutions are used in various fields, such as automotive industry, as cryoprotective agents (CPA) \cite{dashnau2006hydrogen,sieme2016mode,tsuruta1998effects, Pei-Qi_2022,kim2011evaluation,ELLIOTT201774}, skincare products \cite{fluhr2008glycerol,loden2001influence,batt1988changes,bjorklund2013glycerol,rawlings1995effect} and food  or tissues cryopreservation. \cite{hammerstedt1992cryopreservation,keros2005optimizing,storey1998comparison,tada1990cryopreservation,hallak2000cryopreservation,pegg2009principles} 
\textcolor{black}{Among the polyol molecules, glycerol has been demonstrated to be an excellent cryoprotectant. Its activity has been attributed to its ability to break  water hydrogen bonds, thus hindering water molecules  arrangement \cite{cray2013universal}. This happens since glycerol creates hydrogen bonds with its own hydroxyl groups, and has a negative enthalpy of mixing with water.\cite{huemer1994new,marcus2000some} One of the main features of these polyols lies in their amphiphilic nature.  Thus, glycerol  form stable hydrogen bonds with the solvent and, at the same time,  hinder water interaction. Recent studies correlate the enthalpy of mixing of water and glycerol solutions to the number of hydrogen bonds between the two species.\cite{mahanta2023local} 
Due to this feature,  these molecules are used to protect tissues, cells or proteins from  water crystallization and the ensuing damage produced by the growth of ice crystals. }  
\cite{hubalek2003protectants,bhattacharya2018cryoprotectants,storey1998comparison,fahy2015principles}

In the last decades, 
several experiments have been performed 
to study  glycerol-water mixtures, most of them carried out at temperatures above melting \cite{adamenko2006anomalous,takamura2012physical, segur1951viscosity,shankar1994experimental,kwon2000viscosity, nishijima1960diffusion}.
However, the properties of supercooled water-glycerol solutions are not only important for cryopreservation but also provide important insights on the anomalies of water. The complex glass forming behaviour of water-glycerol solutions has attracted much attention.\cite{inaba2007multiple,suzuki2014experimentally,popov2015puzzling,bachler2016glass,alba2022interplay}

Water is one of the simplest liquids in nature but it is, doubtless, the most complex too presenting 19 distinct ice phases when freezing.\cite{chaplin_water}Nevertheless, its complexity is attributed not only to the several solid phases, but also to the anomalies (maxima or minima) that have been reported in the liquid phase for thermodynamic response functions (i.e. isothermal compressibility ($\kappa_{T}$), thermal expansion ($\alpha_{p}$) or specific heat (C$_{p}$)) as well as for dynamic properties. \cite{gallo2016water}

\textcolor{black}{
Suzuki and Mishima \cite{suzuki2014experimentally,suzuki2016effect}, studied a supercooled (and diluted) glycerol-water solution to explore the possible existence of a  liquid-liquid critical point (LLCP) as well as the transition from low density to high density amorphous ices (LDA and HDA respectively).}
\textcolor{black}{
Thereafter, Bachler et al. \cite{bachler2016glass} 
studied polymorphism in water-glycerol solutions at higher concentrations (glycerol mole fraction $\chi_{M}$ $\leq$0.38), demonstrating the complex behaviour of these binary solutions in glassy conditions. }

Experiments on supercooled water-glycerol mixtures have been  complemented with a  theoretical study\cite{jahn2016glass} that not only tried to give a molecular explanation to the anomalies (such as the density maximum)  but also predict other scenarios at temperatures (up to 20 K) and pressures (up to 3000 MPa) that would be experimentally inaccessible.
Another computational study,  by Akinkunmi and coworkers, \cite{akinkunmi2015effects} used TIP3P\cite{jorgensen1983comparison} + R-FF\cite{reiling1996force}, TIP3P + BC-FF\cite{chelli1999glycerol} and TIP4P/2005\cite{abascal2005} + BC-FF force-fields to simulate water-glycerol solutions in a wide range of temperatures and concentrations, even though with a small number of molecules. In this article, the authors observed a density maximum  at low glycerol concentrations that vanished when the concentration increased.
This clearly demonstrated that adding glycerol molecules not only affects ice nucleation but also the  anomalous thermodynamic properties of liquid water. In the same  work\cite{akinkunmi2015effects}, the authors reported the diffusion coefficient for glycerol and water molecules. Both showed a pronounced decrease of the diffusion with glycerol concentrations, independently of the chosen  force-fields.

\textcolor{black}{Despite the widespread use of glycerol-water mixtures, the relevant features of their transport properties still remain to be fully understood, especially in the supercooled regime.}
 The aim of the present work is to unravel the  behaviour of transport properties of water-glycerol mixtures  at low temperatures by means of classical molecular dynamics simulations of all-atom and coarse grained models. 
 We first  study the diffusion of supercooled water (both all-atom and coarse grained)
 and glycerol, finding an unexpected crossover of the diffusion with 1\% and 10\% glycerol concentration. 
  Interestingly, the same crossover is observed when computing the system's shear viscosity for the all-atom system.  
Next, we compute the lifetime of water-water hydrogen bonds and observe a crossover at the same temperature.  
  \textcolor{black}{
To try to explain  this crossover, we compute the temperature of maximum density (TMD) and several structural properties. 
We find that the former is shifted to lower temperatures when increasing the glycerol content, eventually disappearing when the glycerol concentration is above 30\%. 
Whereas the latter show an increase of the  tetra-coordinated water molecule parameter combined with an increase of clustering of glycerol, when supercooling the system. 
}
\section{Simulation Details}
\label{simdetails}

\subsection{All-atom simulations}
We study water-glycerol mixture by means of atomistic simulations as in 
 Ref.\cite{akinkunmi2015effects}, implementing  in the GROMACS (2021) open source molecular dynamics (MD)  package    \cite{gromacs} 
both the TIP4P/2005 water (rigid) \cite{abascal2005} and  the AMBER (BC-FF)  glycerol  (non rigid) \cite{ForceField01,100glyce}. 
\textcolor{black}{The electrostatic long-range interactions were calculated by means of Particle Mesh Ewald method.}
For water-glycerol interactions we used the well known Lorentz-Berthelot mixing rules.\cite{lorentz1881ueber,allen2017computer}
To test the glycerol force field, we first simulate a pure glycerol system, comparing thermodynamic results obtained for pure glycerol to the ones reported in  Ref.\cite{akinkunmi2015effects,100glyce}, such as  the density versus temperature (data not shown). 

\begin{figure}[h!]
    \centering
    a) \includegraphics[width=0.3\columnwidth]{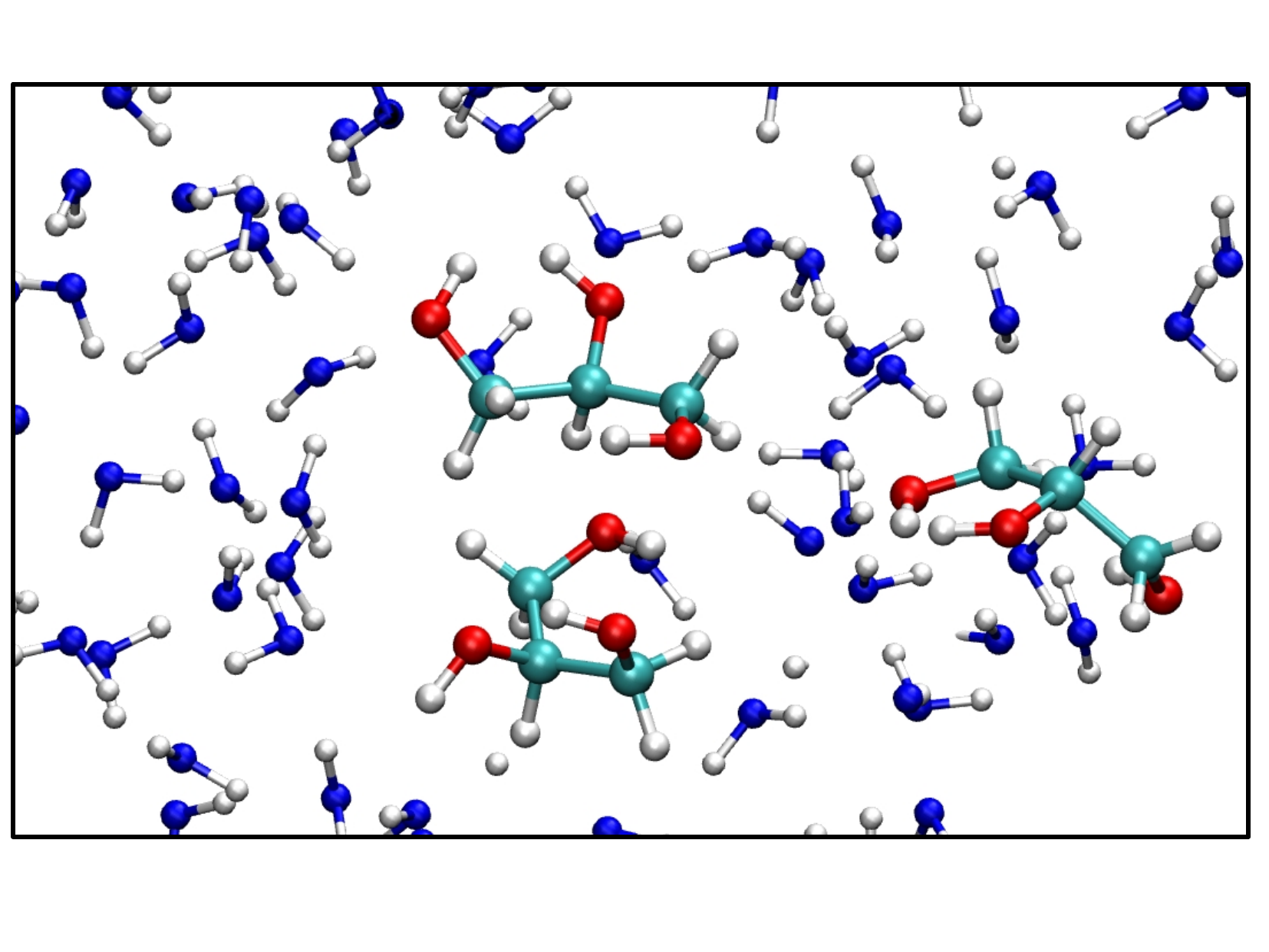} 
         b) \includegraphics[width=0.25\columnwidth]{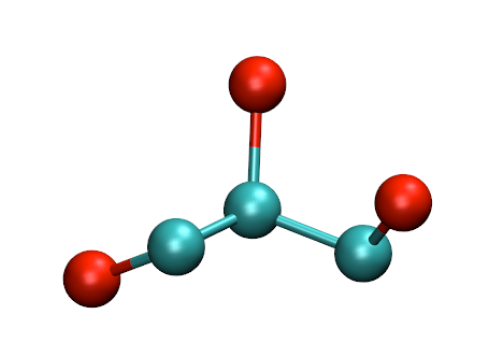}
    \caption{a) Zoom of a snapshot of full-atomistic glycerol molecules ($C_{3}$ $H_{8}$ $O_{3}$ hydrogen in white, oxygen red and carbon in cyan) surrounded by water molecules $H_{2}$$O$ (oxygen in blue and hydrogens in white). b) Snapshot of a six-site united atom coarse-grained glycerol molecule (oxygen in red and carbon in cyan). mW water is not shown. 
    }
    \label{fig:snap}
\end{figure}

Next, we prepare an initial configuration for the all-atom system containing both water and glycerol molecules, 
as the one reported in Figure    \ref{fig:snap} a). 

 To prepare the initial  configuration at several glycerol concentrations, we follow the hereby reported protocol.  
\begin{enumerate}


\item  Having located  a single glycerol molecule in a simulation box, we replicate it until reaching the desired  number of glycerol molecules $N_{gly}$.

\item  We  solvate the glycerol adding water molecules to obtain a given  glycerol (mass) concentration (see Table  \ref{tab:gly-wat}).
\begin{table}[h!]
\tbl{$N_{gly}$ is the number of glycerol molecules,  $N_{wat}$  of water molecules, $\chi_{g}$ the corresponding mass concentration for the all-atom simulations.}
{\begin{tabular}{lcccccccccccccccccc} 
\toprule
    {$N_{gly}$}  &&&&&&  &&& {$N_{wat}$} &&& &&&&& &  $\chi_{g}$ \\ 
 \midrule
      96    && &&&& &&  &     48344  &&& && &&&  & 1\%           \\
 96       && &&&  &&& &  4410     &&  && &&&  &  & 10\%    \\
195    && & && & && &  2332     &&   && & && &   & 30\%    \\ 
195  && & && & && & 992  &  &  &&  &&  &&&       50\%    \\
 \bottomrule
\end{tabular}}
\label{tab:gly-wat}
\end{table}

 As shown   in  Table  \ref{tab:gly-wat}, differently from Ref.\cite{akinkunmi2015effects}, we prepare  water/glycerol mixtures containing a large number of molecules (to improve the statistics and avoid finite size effects).
To directly compare with  Ref.\cite{akinkunmi2015effects}, one should convert mass ($\chi_m = \chi_g$) to molar concentration $\chi_M$
\begin{equation}
    \chi_g = \frac{\chi_{M} \left(\frac{M_{gly}}{M_{wat}}\right)}  {1- \chi_{M} \left(1 - \frac{M_{gly}}{M_{wat}}\right) }
\end{equation}
\textcolor{black}{where $M_{wat}$ is the molecular weight of water (18.01528 g/mol) and  $M_{gly}$ the one of glycerol (92.09382 g/mol).} 
\item  Finally, we  equilibrate the system for  2 ns in an NVT ensemble at relatively high temperature (for instance $T=260$ K).
\end{enumerate}

 Once the water-glycerol mixture is prepared, we simulate the system 
 with a Velocity Verlet algorithm (with time step of 1 fs) in a NPT ensemble   with periodic boundary conditions in all directions. We keep the pressure constant with 
 a Parrinello-Rahman \cite{parrinello1981polymorphic} barostat {\textcolor{black}{ (with a relaxation time of 0.3 ps) } and the temperature constant 
 by means of a velocity rescaling thermostat \cite{grubmuller1991generalized} (with a relaxation time of 0.3 ps). 
Having set the pressure to 1 bar,  we let the system evolve for at least 40 ns in order to equilibrate it at the  density corresponding to the given temperature within this range [210-300]K.
On the other side, production runs have been obtained in a NVT ensemble (at the chosen density and temperature) for at least 650 ns.

\subsection{Coarse-grained simulations}


We perform coarse-grained simulations of glycerol-water solutions using united atom coarse-grained models (UA-CG). These models represent all atoms except hydrogen, and use short-range anisotropic{\cite{molinero2009water}} interactions to model the van der Waals and hydrogen bonding interactions of the mixture. This results in computational efficiency with respect to the all-atom models with long-range electrostatics.

Water is modeled with the monatomic water model mW,\cite{molinero2009water} which represents well the structure, anomalies and phase behavior of water at ambient pressure, as well as producing spontaneous ice crystallization in simulation-accessible times.
Same as TIP4P/Ice, mW presents a continuous transformation from high- to low-density liquid upon cooling at 1 bar.\cite{moore2011structural} 
Different from TIP4P/Ice, mW spontaneously form ice on time scales accessible through simulations and does not have a first order liquid-liquid transition at high pressures.\cite{holten2013nature}

  
\begin{table}[h!]
\tbl{Force field parameters for the interactions between  UA-CG glycerol  and CG mW  water. $\lambda(O_{g}-O{w}) = 24$. glycerol oxygens ($O_{g}$-$O_{g}$). The rest of the Stillinger-Weber parameters for the glycerol-glycerol interactions are same as for mW model. }
{\begin{tabular}{lccccccccccccc}
    \hline
    Pairs &&&&&&& $\epsilon(Kcal/mol)$  &&&&& $\sigma$ (\AA)  \\
    \hline
   $(O_{g}-O_{g})$ &&&&&&& 4.3323 &&&&& 2.2  \\
    \hline
    $(O_{g}-O_{w})$ &&&&&&& 5.85 &&&&& 2.15  \\ 
    \hline
    $(CH-O_{w})$ &&&&&&& 0.17 &&&&& 3.7  \\ 
    \hline
    $(CH_{2}-O_{w})$ &&&&&&& 0.17 &&&&& 3.7  \\ 
    \hline
    $(CH-CH)$ &&&&&&& 0.08 &&&&& 4.043  \\ 
    \hline
    {$(CH-CH_{2})$} &&&&&&&  0.097 &&&&& 4.071  \\
    \hline
  \bottomrule
    \end{tabular}}
    \label{tab:CGparameters}
\end{table}

The fully flexible glycerol molecule is represented by six sites (Fig.\ref{fig:snap} panel b), that reproduces the experimental density, enthalpy of vaporization, and conformational distribution of liquid glycerol modeled. \cite{Katuaprabir2024}  The carbon sites interact through Lennard-Jones interactions and the OH beads through a three-body Stillinger-Weber potential, with the  the parameters shown in Table \ref{tab:CGparameters}. 
Water-glycerol interactions are represented by Lennard-Jones interactions between the C sites and mW and Stillinger-Weber interactions between the OH and mW. 

The parameters, listed in Table \ref{tab:CGparameters}, have been optimized to reproduce  the experimental densities \cite{glycerine1963physical} of the mixture and enthalpy \cite{to1999excess,marcus2000some} of mixing as a function of glycerol concentration, as well as the position of the first peak of the RDF and the number of water neighbors around the glycerol OH groups of the all-atom model of this study  (Figure\ref{fig:enter-label}).
\begin{figure}[h!]
    \centering
    \includegraphics[width=0.5\columnwidth]{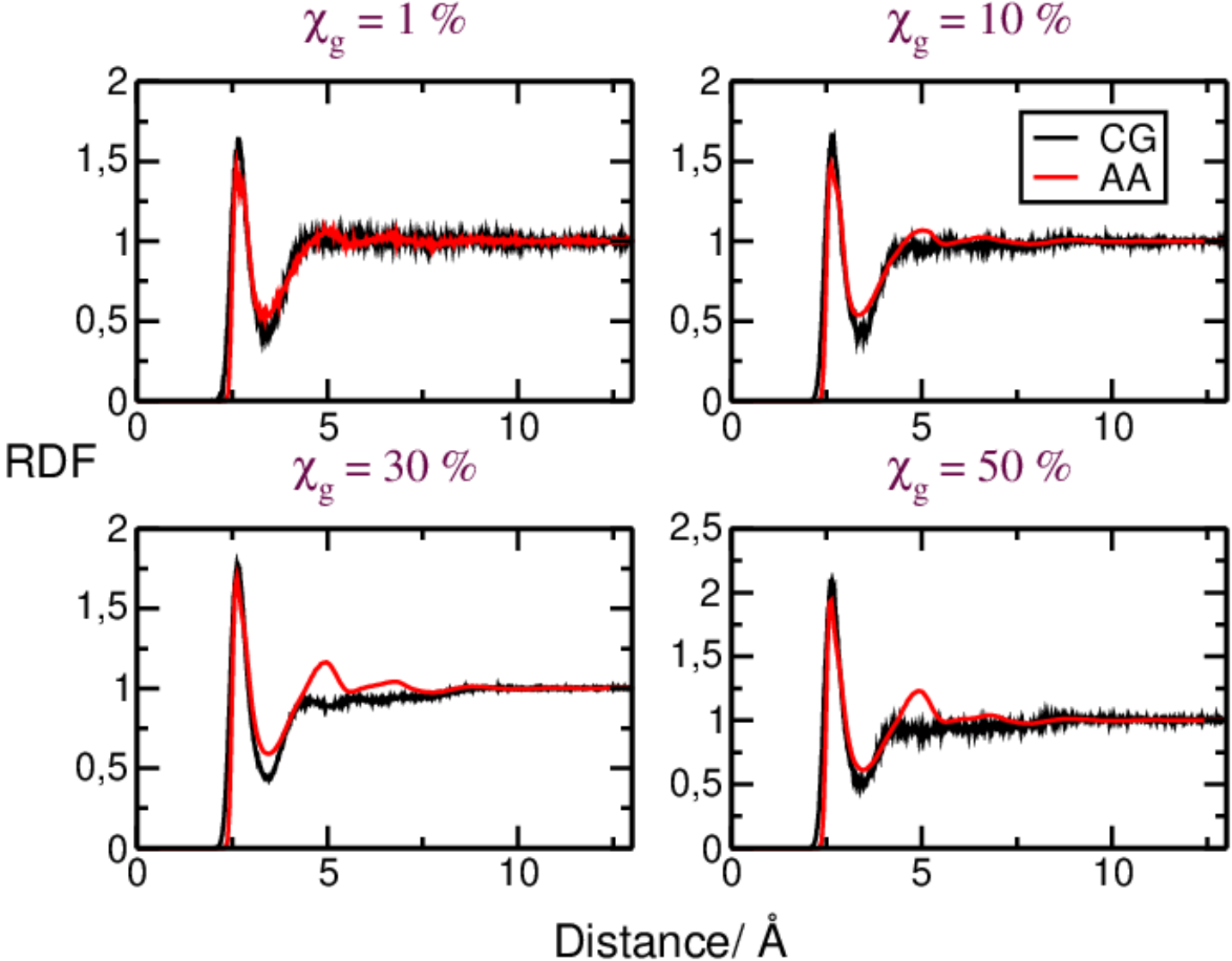}
    \caption{Comparison of radial distribution function between Coarse-grained force-field and all atomistic at T = 298 K at different glycerol concentrations.}
    \label{fig:enter-label}
\end{figure}

The coarse-grained simulation are evolved with LAMMPS.\cite{thompson2022lammps} Input files to run the simulations are provided in the Supporting materials. The equations of motion were integrated with velocity Verlet using 5 fs time steps. The temperature and pressure are controlled with the Nose-Hoover thermostat and barostat with damping times 2.5 and 12.5 ps, respectively. Simulations of water-glycerol mixtures with 1, 10, 30 and 50 percent weight of glycerol are run for 100 ns at temperatures from 298 to 210 K for the parameters reported in Table\ref{tab:CGparameters2}.  

\begin{table}[h!]
\tbl{$N_{gly}$ is the number of glycerol molecules,  $N_{wat}$  of water molecules, $\chi_{g}$ the corresponding mass concentration, T temperature, $\rho$ density and $H^E$ the enthalpy of mixing. }
{    \begin{tabular}{lcccccccccc}
    \hline
    $N_{gly}$ &&& $N_{wat}$ &&& $\chi_{g}$ & T(K)  & $\rho$ (g/$cm^3$) & $H^E$(J/mol) \\
    \hline
     27 &&& 13662 &&& 1\% & 298 &  \\ 
 216 &&& 9942  &&& 10\% & 298 & 1.028 & -331.1 \\
 216 &&& 2576 &&& 30\% & 298 & 1.09 & -273.2 \\ 
 216 &&& 1104 &&& 50\% & 298 & 1.15 & -315.6 \\
  \bottomrule
    \end{tabular}}
    \label{tab:CGparameters2}
\end{table}

    

\subsection{Analysis tools}
In our work we compute transport properties such as the diffusion coefficient and the shear viscosity. 
The diffusion coefficient  is computed via the long time behaviour of the mean square displacement of the centre of mass of either oxygen-glycerol or oxygen-water. 
\begin{equation}
    D =\frac{\left\langle  |\textbf{r}(t)-\textbf{r}(0)|^2  \right\rangle}{6 t}
    \label{eq:MSD}
\end{equation}
where $\textbf{r}$ is the position of the oxygen (either belonging to the water molecule or to the glycerol molecule). 
The shear viscosity is computed via the Green-Kubo relation\cite{GreenKubo} 
\begin{equation}
\eta = \frac{V}{k_B T} \int_0^\infty \left \langle P_{\alpha \beta}(0)\, P_{\alpha \beta}(t)\right \rangle \, dt
\label{eq:eta}
\end{equation}
where $V$ is the volume of the system box, $k_B$ is the Boltzmann constant, $T$ is the absolute temperature and $P_{\alpha \beta}(t)$ are the component for the stress tensor at time $t$. The average runs over the P$_{xy}$,P$_{xz}$ and P$_{yz}$ components of  the pressure tensor. 

Using the results for diffusion and viscosity, we can test the SE relation which writes:
Stokes-Einstein relation for both glycerol and water  \cite{StokeEinstein} 
\begin{equation}
\frac{D \eta}{T} = constant
\label{eq:eta_eins}
\end{equation}
where $D$ is the diffusion constant previously computed (for either oxygen-glycerol or oxygen-water) and $\eta$ the viscosity.

A glycerol molecule contains three oxygen atoms. Typical conformations of a glycerol molecule are shown in Figure \ref{fig:snap1}.
\begin{figure}[h!]
    \centering
   \includegraphics[width=0.33\columnwidth]{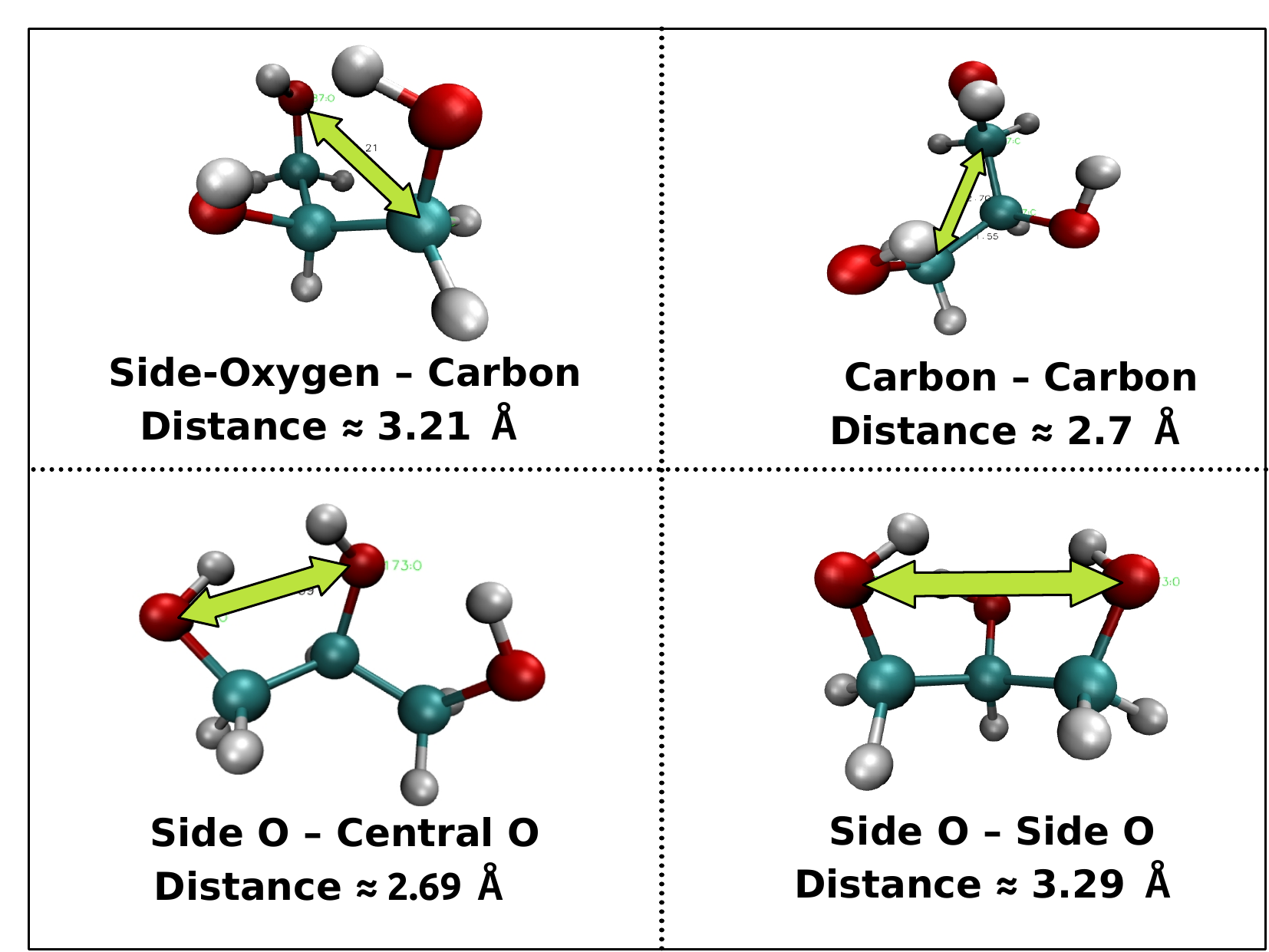}
    \caption{
     Structure of a glycerol molecule $C_{3}$ $H_{8}$ $O_{3}$: hydrogen in white, oxygen red and carbon in cyan.      Intramolecular distances are reported between pairs of atoms.}
    \label{fig:snap1}
\end{figure}

It is important to mention that when computing transport and structural properties of the oxygen of the glycerol molecule,  we cannot distinguish between the central and the outer oxygen atoms.


We estimate the temperature of density maximum computing the density as a function of temperature for solution of mass concentration  $ \chi_g = 1\%, 10 \%, 30\%, 50\% $ and compare to the pure water case.

In order to understand the microscopic behaviour of the supercooled mixture, we compute 
1) the hydrogen bond life time. 
2) the ratio of tetra-coordinated water molecules (P4), 
3) the radial distribution function, 4) the number of hydrogen bonds.

The radial distribution function ($g(r)$ or RDF) gives information on the structure of the liquid molecules surrounding a central one. 
To calculate the RDF,  we have used OVITO \cite{ovito}, and GROMACS\cite{gromacs}. 
To calculate the number of neighbours within each coordination shell, we have integrated the RDF.

To calculate the P4 parameter, we have used the coordination analysis command in OVITO \cite{ovito} and  computed the amount of four-coordinated water molecules in each frame, averaging over the number of water molecules and number of frames. 
It is important to mention that for this analysis we did not consider the glycerol molecules surrounding the water ones, being P4 the fraction of water molecules that are tetra-coordinated by other water molecules.

Using GROMACS, we have calculated the water-water and water-glycerol hydrogen bonds. These bonds are formed when the donor-acceptor distance is $\leq$ 3.5 $~$\AA$~$ and the angle between the OH of the donor and the O of the acceptor is 30º. We calculate the total number of hydrogen bonds averaging over the total number of water (or glycerol) molecules and over the number of frames.

\textcolor{black}{As in \cite{michaud2011mdanalysis,gowers2016proceedings,smith2019interaction}, the hydrogen bonds lifetime $\tau$ is estimated by fitting the decay of the autocorrelation function (ACF) to a double exponential curve:}

\begin{equation}
C_{HB}(t) \approx A_{0} exp(-t/\tau_{1}) + A_{1} exp(-t/\tau_{2})
\label{acfexp}
\end{equation}

\textcolor{black}{
where $\tau_{1}$ and $\tau_{2}$ are two time constants  (a short-timescale process and  a longer one),  and $A_{0}$ and $A_{1}$ weight the  relevance of the short- and longer-timescale processes in the  autocorrelation curve. 
Once we have calculated $A_{0}$, $A_{1}$, $\tau_{1}$ and $\tau_{2}$, $\tau$ can be written as:}

\begin{equation}
\tau (ps) = A_{0} \tau_{1} + A_{1} \tau_{2}
\end{equation}

\textcolor{black}{The hydrogen bonds lifetime ($\tau$) is calculated by means of the MDAnalysis python   tool\cite{michaud2011mdanalysis,gowers2016proceedings,smith2019interaction} that follows the Luzar and Chandler\cite{luzar1996hydrogen} criterion. 
$\tau $ is averaged over 100 frames for each run.}

To calculate the error bars in each measurement, we computed a block average in all cases. Only for the viscosity computed at $\chi_{g}$ = 30\%, we have calculated the error bars using the viscosity values obtained for each component of the pressure tensor.

\section{Results}



We start by characterising the transport properties of the mixture,  
computing the diffusion coefficient of the centre of mass of the glycerol molecules and of the centre of mass of the water molecules, as a function of temperature. 
Figure \ref{fig:Diffusion} reports the glycerol diffusion coefficient $D_{gly}$ (left panel) and the water diffusion coefficient  $D_{wat}$ (right panel) of the all-atom (AA) model at temperatures ranging from T = 210~K to T = 298~K and for glycerol concentrations $\chi_g$ = 1\% (blue symbols),10\% (red symbols), 30\% (black symbols), 50\% (magenta symbols) as well as for pure water \cite{montero2018viscosity} (green line with yellow symbols).



As expected, the diffusion coefficients of  glycerol (left panel) and water (right panel) decrease monotonously with decreasing temperature.  
The left panel of Figure \ref{fig:Diffusion} shows the glycerol  diffusion coefficient.  The diffusion coefficient decreases with increasing glycerol concentration, with the lowest diffusion displayed by the  50\% mixture (magenta curve).   
 Analysing the glycerol diffusion in the $\chi_g$ = 10\%, 30\% and 50\% mixtures, a monotonic decrease is observed with decreasing temperature.
Interestingly, at temperatures lower than 240 K,  
 we detect a crossover between the $\chi_g$ = 1\% and the $\chi_g$ = 10\% diffusion curves.  
\begin{figure}[h!]
    \includegraphics[width=0.45\columnwidth]{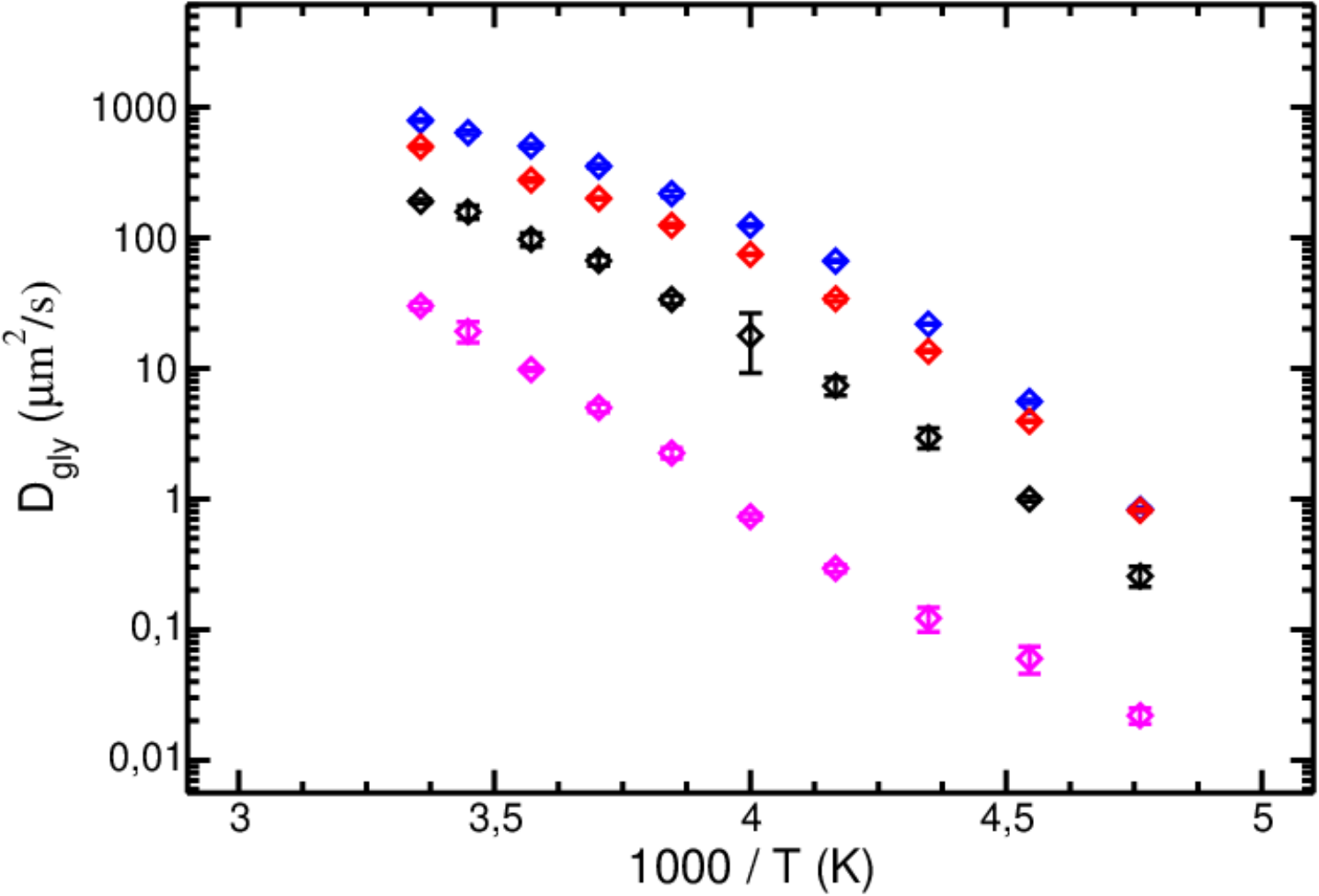}
    \includegraphics[width=0.45\columnwidth]{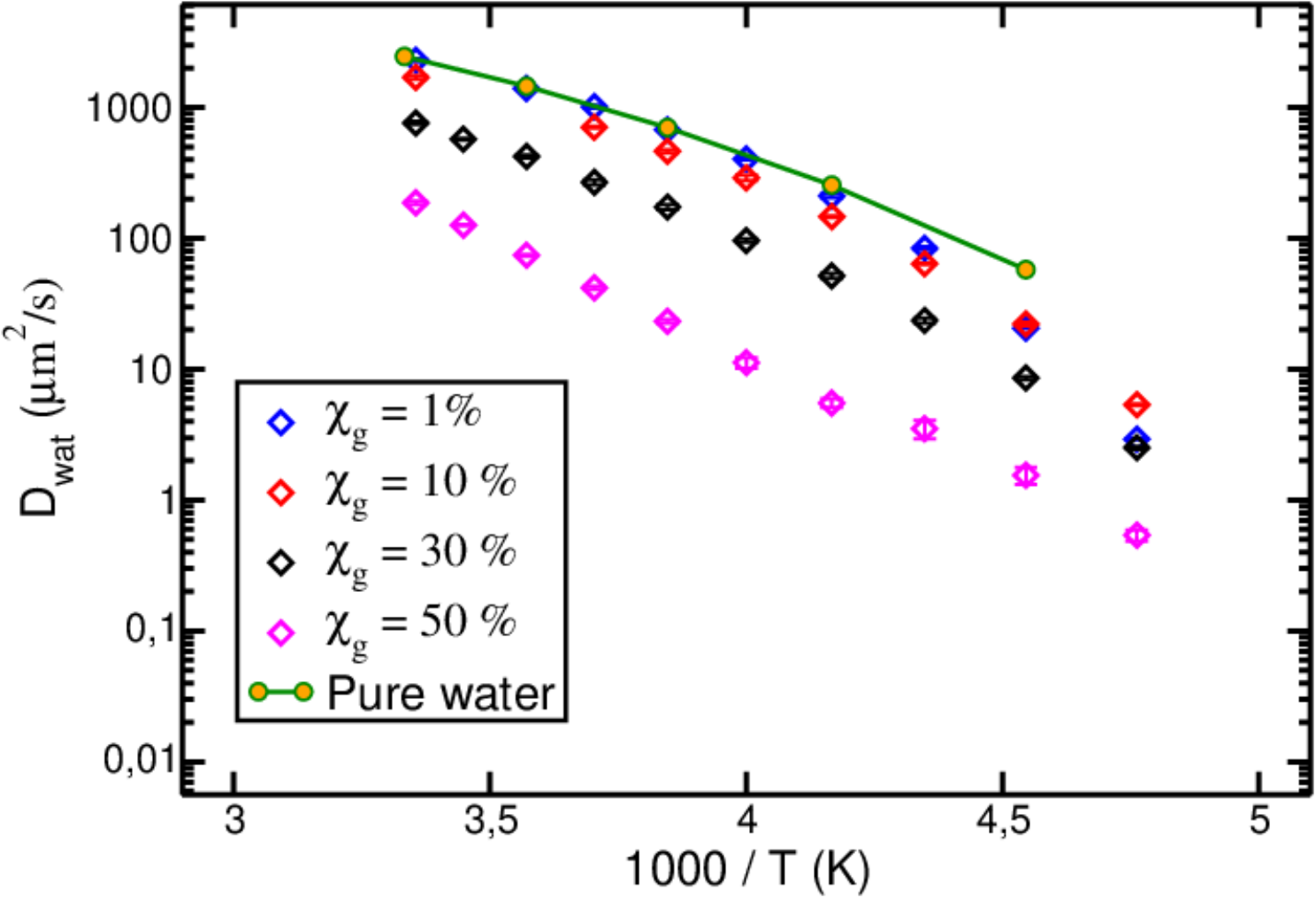}
       \caption{Diffusion coefficient of  glycerol (left) and water (right)  at different  glycerol concentrations $\chi_{g}$: 1\% (blue symbols), 10\% (red symbols), 30\% (black symbols) and 50\% (purple symbols) ).  Water diffusion is reported in the right panel from Ref.\cite{montero2018viscosity} at $\rho$= 999.26 kg$\cdot$m$^{-3}$}
    \label{fig:Diffusion}
\end{figure}

When studying the water diffusion for the all-atom model (right panel of Figure \ref{fig:Diffusion}), 
we find that  diffusion in the most dilute solution, $\chi_g$ = 1\%  (blue curve) is indistinguishable from the one in pure water (green line) at temperatures above T $\geq$ 240 K (blue symbols). 
However, as in the glycerol case, at temperatures lower than 240 K  water diffusion shows a  pronounced decrease (blue diamonds) and at T = 220~K the diffusion of water molecules in the $\chi_g$ = 1 \% solution is the same as in the 10 \% (red symbols) mixture.
This anomalous decrease of the water diffusion continues until T = 210~K where the diffusion of water is higher in the 10\% mixture  than in the 1\% mixture. 

Figure   \ref{fig:diff_CG} reports the diffusion coefficient for water simulated with the coarse grained model for different glycerol concentrations.
 \begin{figure}[h!]
    \centering
    \includegraphics[width=0.45\columnwidth]{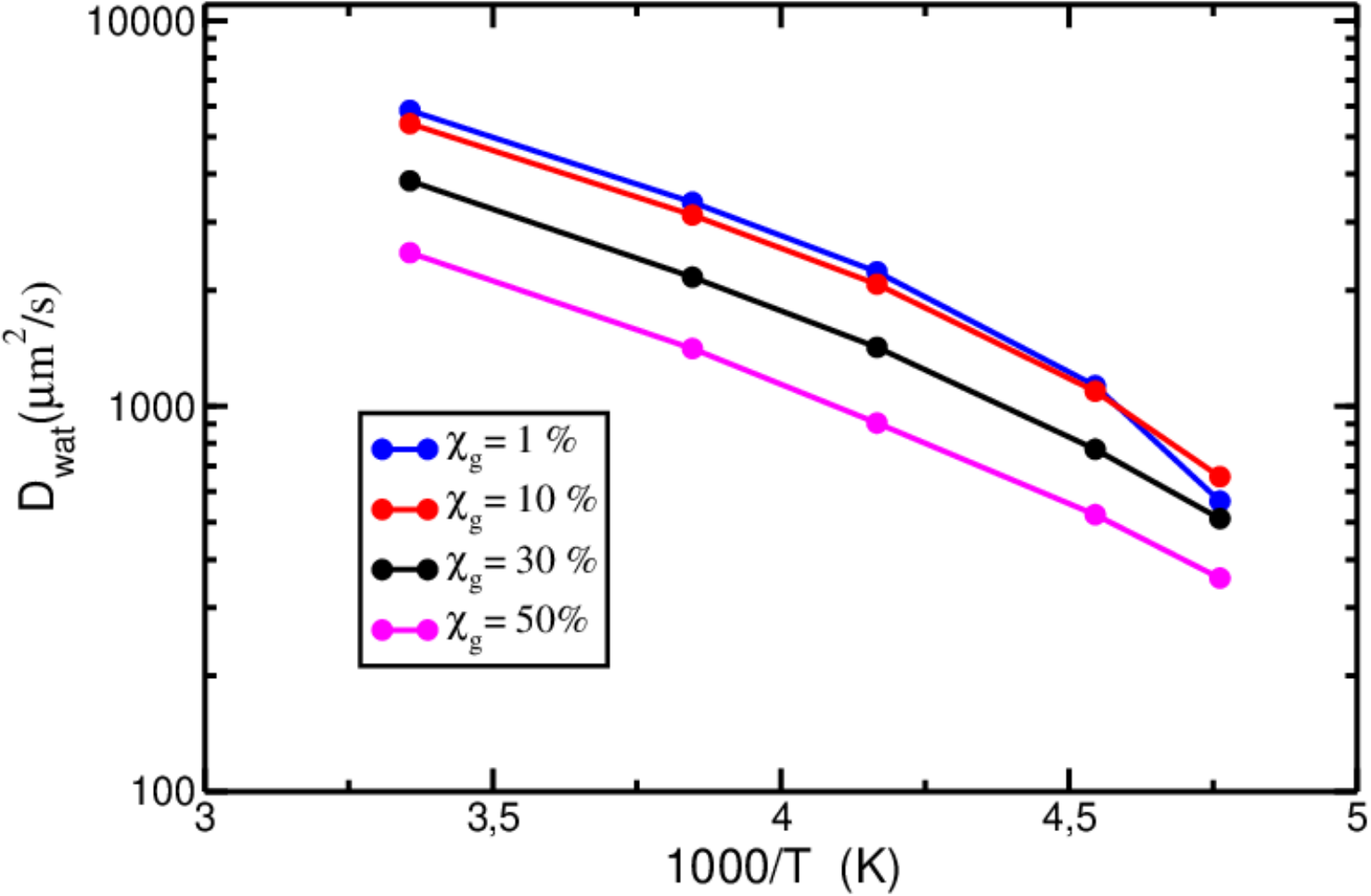}
    \caption{Diffusion coefficient for water molecule calculated for mW.}
    \label{fig:diff_CG}
\end{figure}

The diffusion at $\chi_g$ = 1 \% solution shows the same crossover than the water diffusion for the all-atom simulations. The agreement in the results with the AA and CG models supports that at very low temperatures the diffusion of water is not monotonous with glycerol concentration.

To understand whether the unexpected crossover on the dynamics observed for water and glycerol at low  concentration is only present in the diffusion coefficient, we compute for the AA model different transport properties: hydrogen bond dynamics and the shear viscosity.

After observing the anomalous crossover in the diffusion coefficient, we wanted to verify whether the same event occurs the hydrogen bonds lifetime. We compute the lifetime, as explained in section \ref{simdetails}, by fitting the decay of the autocorrelation function to a double exponential function (see equation \ref{acfexp}). 
This analysis has been carried out for pure water and those concentrations at which we observed the crossover in the dynamic properties, i.e. $\chi_{g}$=1 \% and  $\chi_{g}$=10\%.  In figure \ref{lifetimewaterwater} we represent the values of $\tau$ (in ps) calculated for water-water interaction. 

\begin{figure}[h!]
    \centering
    \includegraphics[width=0.45\columnwidth]{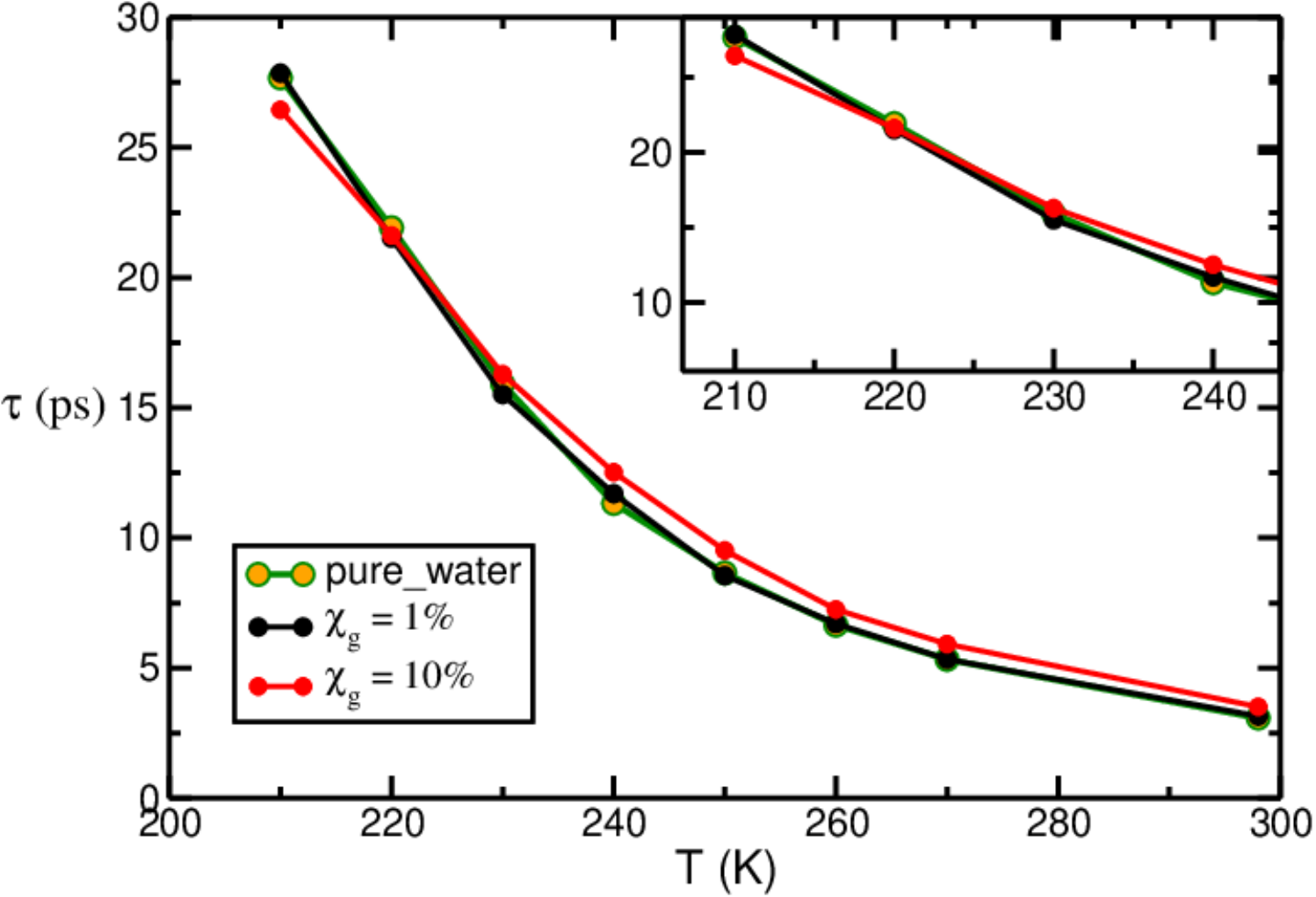}
    \caption{Lifetime of water-water hydrogen bonds computed for pure water (green) $\chi_{g}$=1\% (black), $\chi_{g}$=10\% (red). For clarification we have included an inset for the lowest temperatures.}
    \label{lifetimewaterwater}
\end{figure}

The three systems show a very similar behavior along all temperatures, starting from T = 298 K the lifetime is $\approx$ 3 ps showing a good agreement with previous results \cite{antipova2013hydrogen,guardia2006hydrogen}. Martí et al. \cite{guardia2006hydrogen} reported as "intermitent hydrogen bonds lifetime" for the model SCP/E $C^{int}_{HB}$ = 3.9 ps. Later on, Antipova et.al \cite{antipova2013hydrogen} calculated values for the lifetime of the hydrogen bond in pure water at normal conditions, using different water models and  approximations. For TIP4P model they reported 3 ps using a similar fit as in equation \ref{acfexp}.

Our results, indicate that when decreasing the temperature, the lifetime increases dramatically up to reach $\approx$ 27 ps at T = 210 K. Interestingly there is a crossover at T = 220 K where the lifetime for $\chi_{g}$=10\% crosses the curves for pure water and $\chi_{g}$=1\%. 
}

\begin{figure}[h!]
    \centering
    \includegraphics[width=0.45\columnwidth]{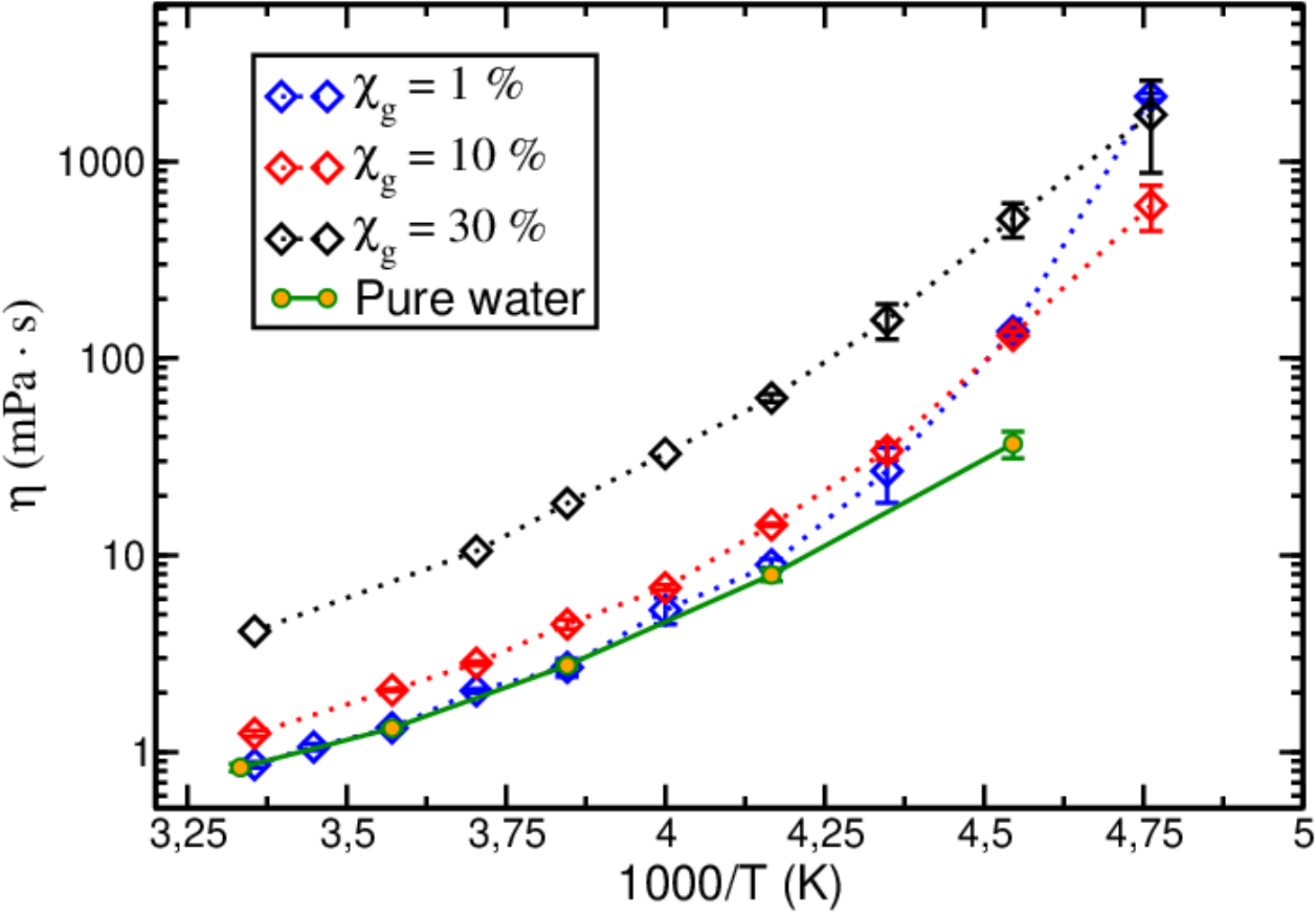}
    \caption{Glycerol-water mixture viscosity   for different glycerol concentrations 
    (1\% in blue, 10\% in red, 30\% in black). The results for pure water are reported in green from Ref.\cite{montero2018viscosity} at $\rho$= 999.26 kg$\cdot$m$^{-3}$.}
    \label{fig:Viscosity}
\end{figure}

\textcolor{black}{Next, we focus on the shear viscosity. Given that the crossover in diffusion coefficient is only present at low glycerol concentration, we have decided not to compute the shear viscosity at the highest concentration of   $\chi_g$ = 50 \%.}
Figure \ref{fig:Viscosity} reports the shear viscosity computed for 
  $\chi_g$ = 1 \%, 10 \% and 30 \%  glycerol concentrations, as compared to the viscosity of pure water (in green).

As expected, Figure \ref{fig:Viscosity} shows that viscosity increases with increasing glycerol concentration and  decreasing temperatures. 
Interestingly, at the lowest temperatures (around 220 K) we find  the same crossover  between the  1\% and the 
 10\% glycerol concentration curves unveiled by the diffusion coefficients:  the 1\% concentration solution has higher viscosity than the 10\%   solution.
All numerical values reported in figure \ref{fig:Viscosity} have been plotted in the following table  \ref{tab:viscosity}.
\begin{table}[h!]
\tbl{Viscosity computed by means of Green-Kubo relation for 1\%, 10\% and 30\% concentrations. The values for pure water were obtained from \cite{montero2018viscosity}.*Interpolated values using the data from the same article. }
{\begin{tabular}{lcccccccccccc} 
\toprule
T(K)  &&&   Water  &&&   1\% &&&  10\% &&& 30\% 
    \\ 
 \midrule
298 &&&0.87*&&& 0.86 &&& 1.24&&& 4.11  \\
270 &&& 2.05* &&& 2.05 &&& 2.83 &&& 10.48 \\
260 &&& 2.75 &&&2.69 &&& 4.46 &&& 18.31  \\
250 &&& 4.15* &&&5.26 &&& 6.84 &&& 32.86  \\ 
240 &&& 7.94 &&&8.95 &&& 14.24 &&& 62.86\\
230 &&& 17.52*  &&&26.78 &&& 33.94 &&& 156.49 \\
220 && & 36.80 &&&137.31 &&& 130.76&&& 510.68 \\
210&& & - &&&2136.16&&&  598.08 &&& 1725.60 \\
 \bottomrule
\end{tabular}}
\label{tab:viscosity}
\end{table}

\begin{figure}  [h!] 
\includegraphics[width=0.45\columnwidth]{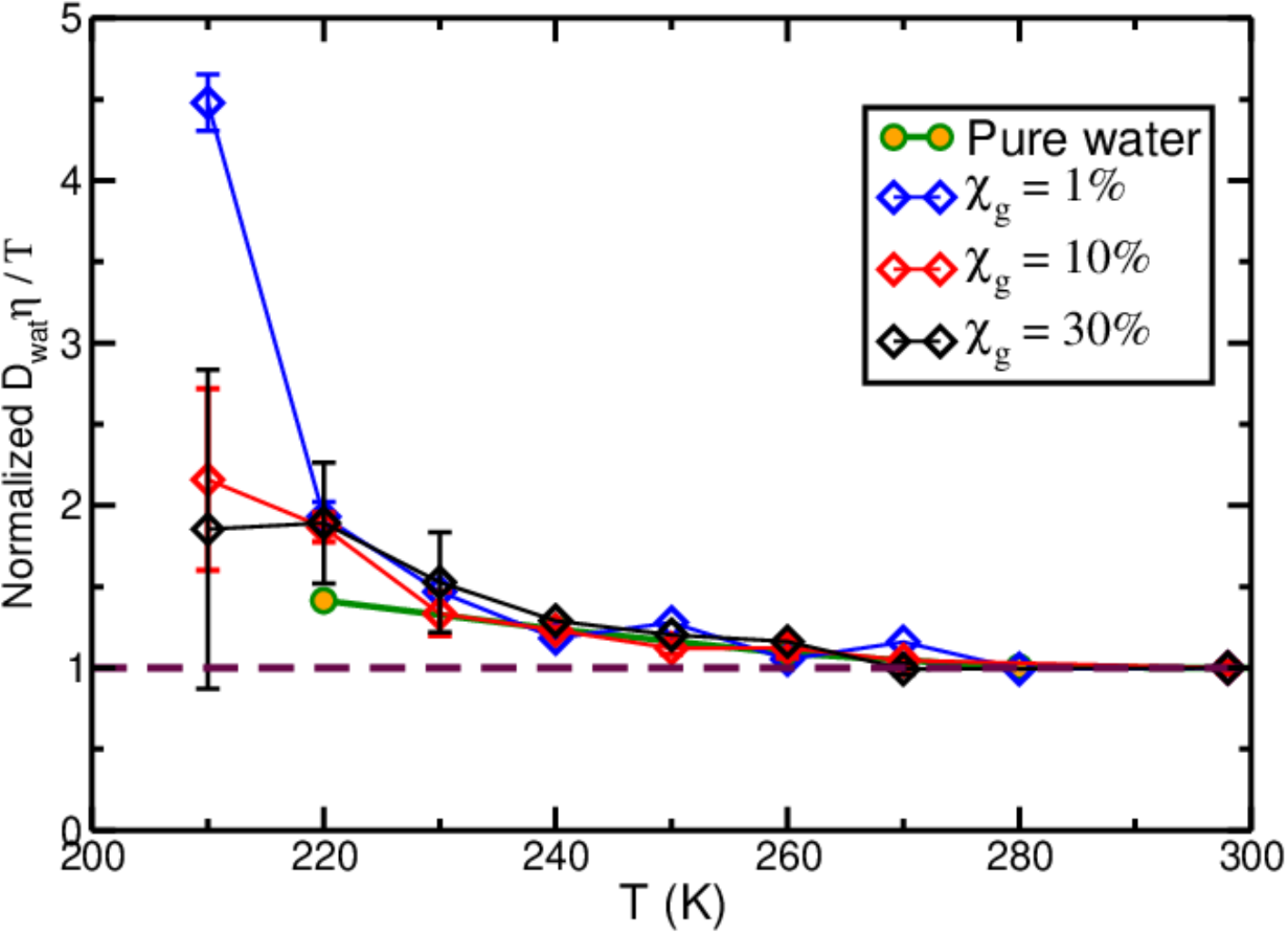}
 \includegraphics[width=0.45\columnwidth]{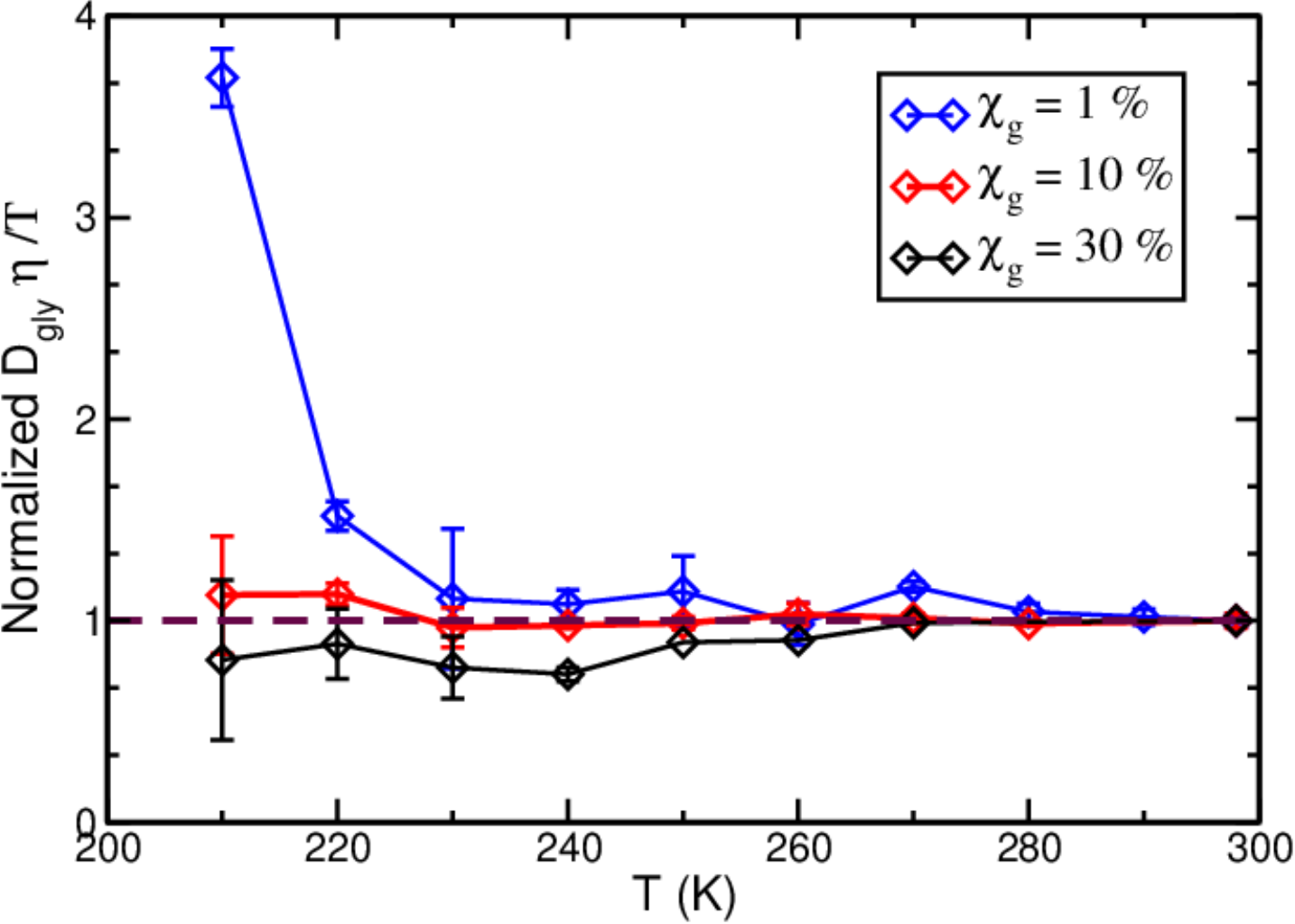}
    \caption{Normalised Stokes Relation for water molecules (left) and glycerol molecules (right) evaluated for all the concentrations as well as for pure water. Pure water  from ref\cite{montero2018viscosity}.}
    \label{fig:SE_watgly}
\end{figure}

Knowing the diffusion and the viscosity,  we use the results of the AA simulations to  assess the validity of the Stokes-Einstein relation (eq. \ref{eq:eta_eins}) for water (Figure \ref{fig:SE_watgly}-left panel) and glycerol (Figure \ref{fig:SE_watgly}-right panel). We find that the mixture follows the Stokes-Einstein relation at high  temperature for both water and glycerol. However, when the temperature is below 220K, the behaviour is more complex. In the case of 
water (left-panel), the normalised ratio of diffusion and viscosity of the $X_g$=1\% solution significantly departs from 1 when the temperature is ~220 K or lower.  
This is the same temperature where the crossover has been observed in the 
  diffusion coefficient.
The same behaviour is observed for the 1\%  glycerol solution (blue curve) in the right-panel, when we focus on the glycerol instead of water.

Even the magnitude of the violation of SE and the temperature of the onset of the deviation are the same for water and glycerol in the most diluted solutions (blue curves in both panels). Interestingly, the 30\% solution on the right panel present a slightly negative deviation from Stokes-Einstein, indicating that the glycerol molecules move more sluggishly than expected from the viscosity. This negative deviation may be associated to the development of dynamic clusters of glycerol in the concentrated solutions (see Figure \ref{fig:clusters} in Appendix). 


When dealing with pure water, the authors of Ref.\cite{montero2018viscosity} suggested that the Stokes-Einstein violation observed in an all-atom water simulations at density 999.26 kg$\cdot$ m$^{-3}$
below 240K is  related to the existence of a line of maxima in heat capacity (Widom line) emanating from the Liquid-Liquid critical point (LLCP) in supercooled water.  TIP4P/2005 has a liquid-liquid transition with a LLCP at 135 MPa and 193 K \cite{abascal2010widom} and the maximum in heat capacity at 1 bar occurs at 245 K.\cite{biddle2017two} 

The coarse grained mW model for pure water has a  maximum in density at 250 K and for the response functions at 201 K at 1 bar,\cite{molinero2009water,moore2011structural} but does not have a first order liquid-liquid transition. \cite{limmer2011putative, holten2013nature} The maximum in response functions at the Widom line is associated to a structural transformation of water towards a four-coordinated liquid.\cite{moore2011structural}
To understand whether the anomalies of water are related to the crossover in the dynamics of the water-glycerol mixture, we compute the temperature of maximum density of the solutions as a function of concentration for both all-atonm and coarse grained models.

\begin{figure}[h!]
    \centering
    \includegraphics[width=0.43\columnwidth]{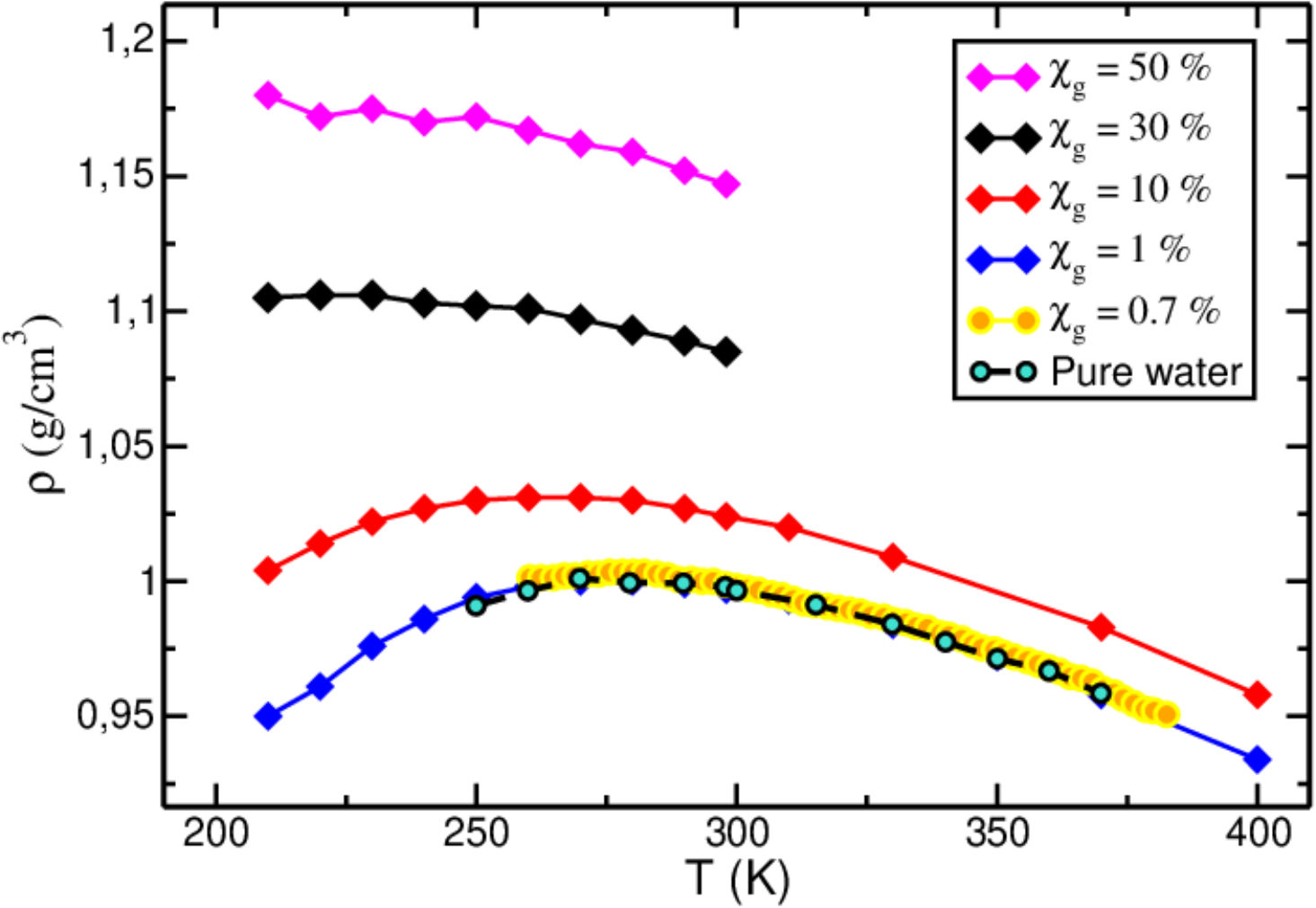}
        \includegraphics[width=0.42\columnwidth]{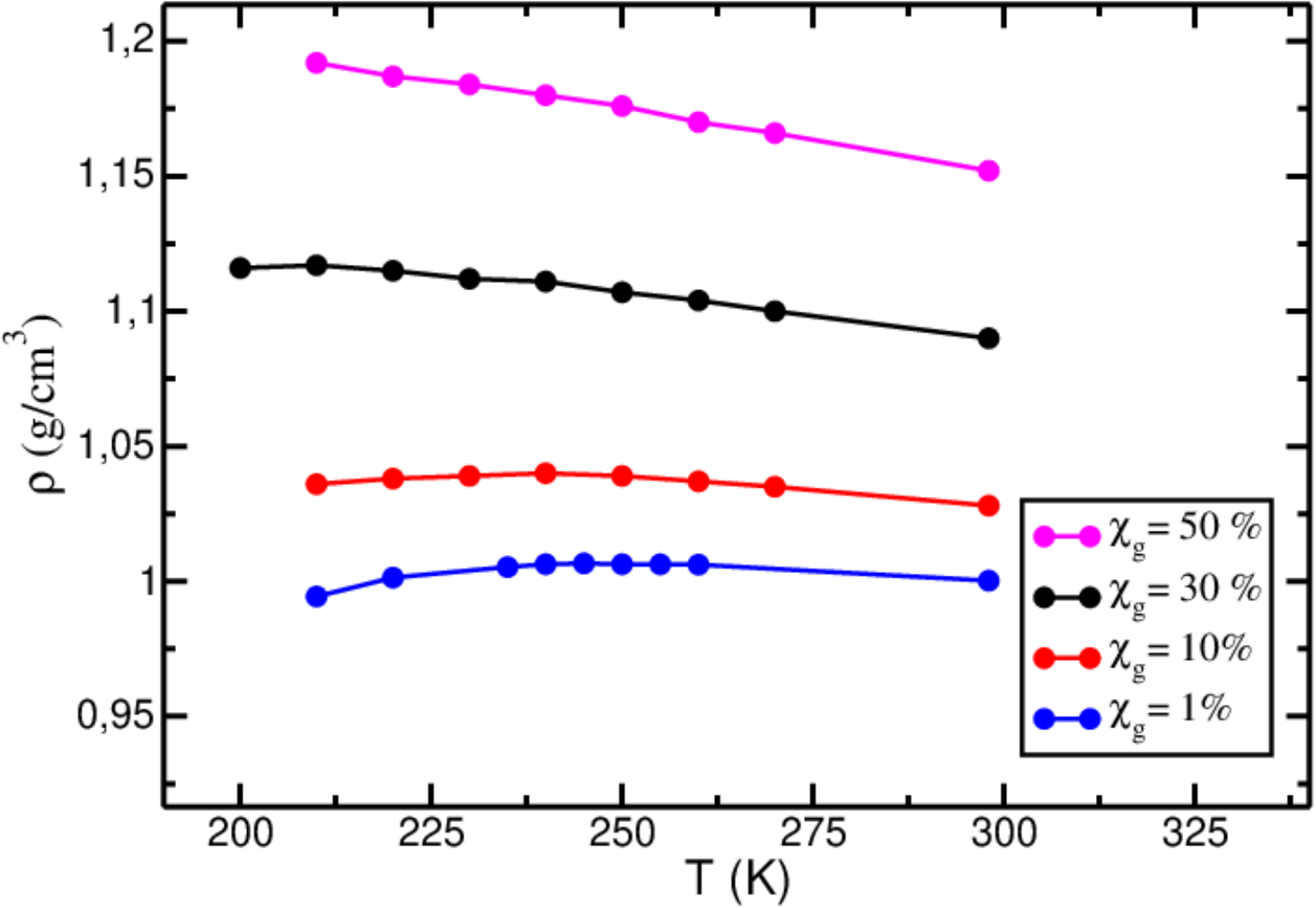}
    \caption{Density vs Temperature for different  concentrations: 1\% in blue, 10\% in red, 30\% in black. $\chi_g$=0.7\% (in yellow) and  pure water (in green) are  from Ref \cite{akinkunmi2015effects}. Left-hand side: full atomistic. Right-hand side: coarse grained.}
    \label{fig:Density}
\end{figure}

 Figure \ref{fig:Density}  presents the density versus temperature (T)  computed with the AA model (left-panel) and CG model (right-panel) for mixtures at all glycerol concentrations. 
   Figure  \ref{fig:Density} (left)presents the density vs temperature for the solutions modeled with the AA potential  together with the results at the lowest concentration from ref\cite{akinkunmi2015effects} (in yellow) and those   obtained for pure water from ref.\cite{montero2018viscosity} (cyan and black circles). 
  Figure  \ref{fig:Density} (right) presents the density vs temperature for the solutions modeled with the CG potential. 

  In both models, the TMD shifts to lower temperatures when the glycerol concentration increases,  until it vanishes for $\chi_g$ = 30\%. The disappearence of the TMD with increasing solute concentration is  in agreement with the results reported in \cite{akinkunmi2015effects} for water glycerol mixtures and with the work by Ref.\cite{Gallo2011} on supercooled salty aqueous solutions. 

Given that the behaviour of the TMD  for 1\% water-glycerol mixture is the same as that of the pure water one, we  suggest that  in the water-glycerol mixture  the Stokes-Einstein violation could be  related to the structural transformation into a four-coordinated low-density liquid.

For both the AA and CG models the dynamic crossover occurs well below the TMD of pure water and the 1\% solution. Interestingly, the crossover temperature for the AA model occurs below the 245 K Widom temperature of pure TIP4P/2005,\cite{abascal2010widom} while for the CG model is above the 201 K Widom temperature of pure mW.\cite{moore2011structural} 
However, differently from pure water, the mixture contains glycerol molecules that interact with water:  even though this does not affect the position of the density maximum of the solution, which is the same as for pure water, the presence of 1\% glycerol decreases the mobility of water in the deeply supercooled region.

\begin{figure}[h!]
    \centering
 \includegraphics[width=0.45\columnwidth]{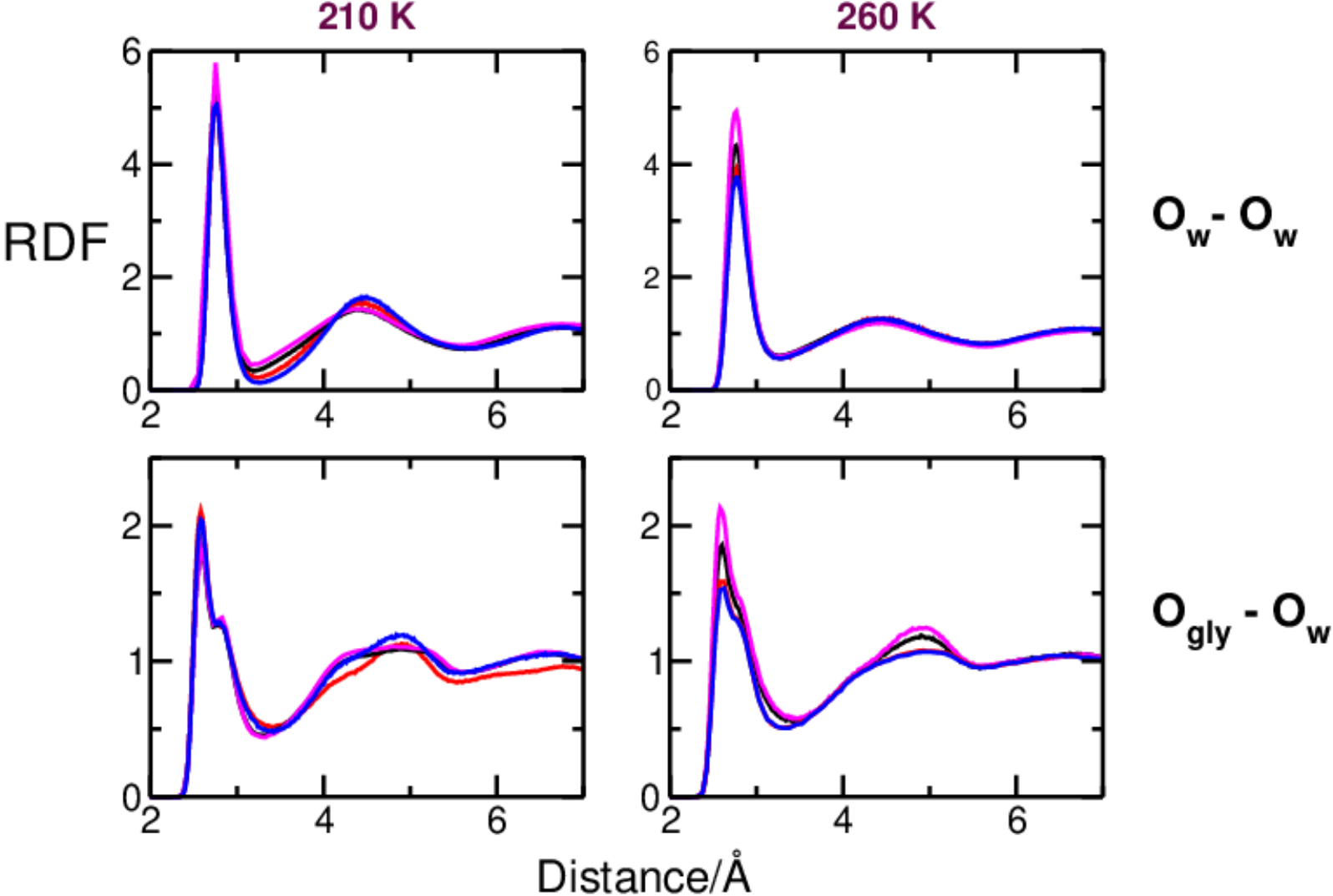}
           \caption{Radial distribution function for O$_{water}$ - O$_{water}$ and O$_{water}$ - O$_{gly}$ ,  at T = 210 K (left panels) and  T = 260 K (right panels) for all the concentrations: $\chi_{g}$ = 1$\%$ (blue) , $\chi_{g}$= 10 $\%$ (red), $\chi_{g}$ = 30$\%$ and $\chi_{g}$ = 50$\%$ (magenta). }
    \label{fig:rdfall} 
\end{figure}

To better characterise the dynamic crossover, we also study several microscopic properties. 
We start by computing the radial distribution function (RDF) between the (water)oxygen-(water)oxygen and  (glycerol)oxygen - (water)oxygen molecules (Figure \ref{fig:rdfall}) for the AA model.  
To unravel potential differences, we compute the RDF at high temperature (260 K) and below the dynamical crossover (210 K).

Comparing the RDF plotted in the left panels (low temperature) to those in the right panels (high temperature), we conclude that the differences are rather subtle.
It might be noticeable the second peak at T = 210 K where the intensity of the second peak of glycerol-water interactions at $\chi_{g}$= 10 $\%$ looks less pronounced than the others, which would suggest a less structured behaviour.

\begin{figure}[h!]
    \centering
        \includegraphics[width=0.4\columnwidth]{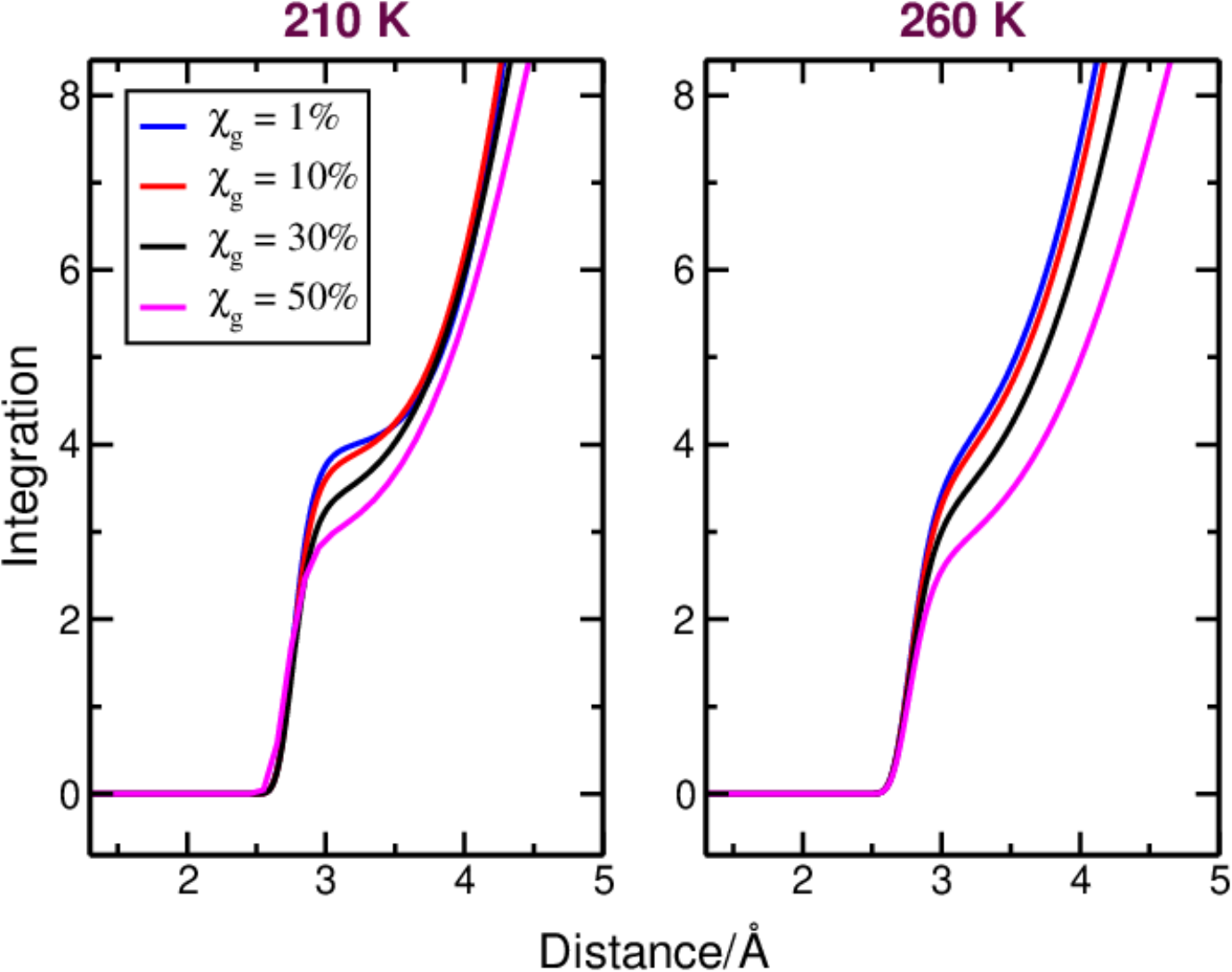}
    \caption{Integration of radial distribution function calculated for water-water at T = 210~K and T = 260~K for concentrations $\chi_{g}$ = 1$\%$ (blue) and $\chi_{g}$ = 10$\%$ (red), $\chi_{g}$ = 30$\%$ and $\chi_{g}$ = 50$\%$ (magenta).}
   \label{fig:neighbwatwat} 
\end{figure}

In Figure \ref{fig:neighbwatwat} we represent the integration of the RDF for the first coordination shell of water-water interaction at 210 K and 260 K. 

It can be observed how, for both temperatures, the number of water-water neighbors decrease with glycerol concentration. It is also noteworthy that, at the lowest temperature, there is a clear plateau for $\chi_{g}$ = 1 \%. This plateau would suggest a more structured behaviour in the vicinity of the water molecules which is less pronounced at higher concentrations.

\begin{figure}[h!]
\centering
\includegraphics[width=0.4\columnwidth]{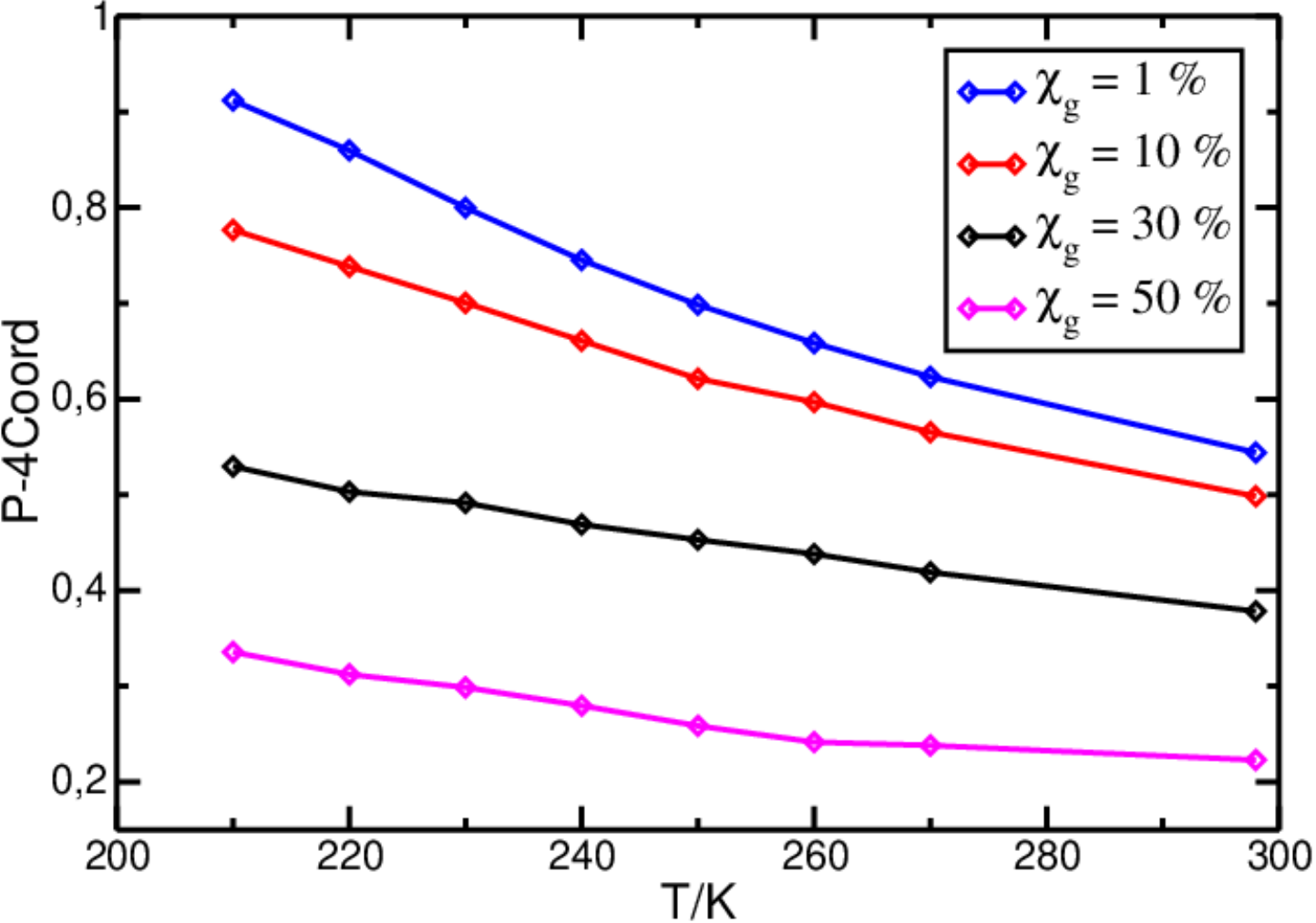}
\caption{Four-coordinated water molecules as a function of temperature (K) for all the systems studied. All values of P4 are computed with respect to the value of  P4 at T = 298 K. }
\label{fig:distances} 
\end{figure}

Next, we compute the number of 4-coordinated water molecules, \textcolor{black}{normalized with respect to the total number of water molecules in the system.} (Figure \ref{fig:distances}). 

The number of 4-coordinated water molecules decreases when the temperature increases and also when increasing the glycerol concentration. However,  no anomalous behaviour is detected at low $\chi_g$ that could justify the crossover in the dynamical properties.

A possible explanation for the dynamical crossover might be due to the interplay between the decrease in mobility due to increase in the fraction of tetrahedral coordinated  water molecules and the number of  glycerol molecules. 
We interpret that the decrease in mobility arising from the development of larger fraction of four coordinated water molecules at higher temperatures in the 1\% solution outcompetes the increase in mobility due to strong water-glycerol and glycerol-glycerol interactions in the 10\% solution.
The crossover between 1\% and 10\% suggests that at that temperature both contributions are comparable.



\begin{figure}[h!]
\centering
     \includegraphics[width=0.45\columnwidth]{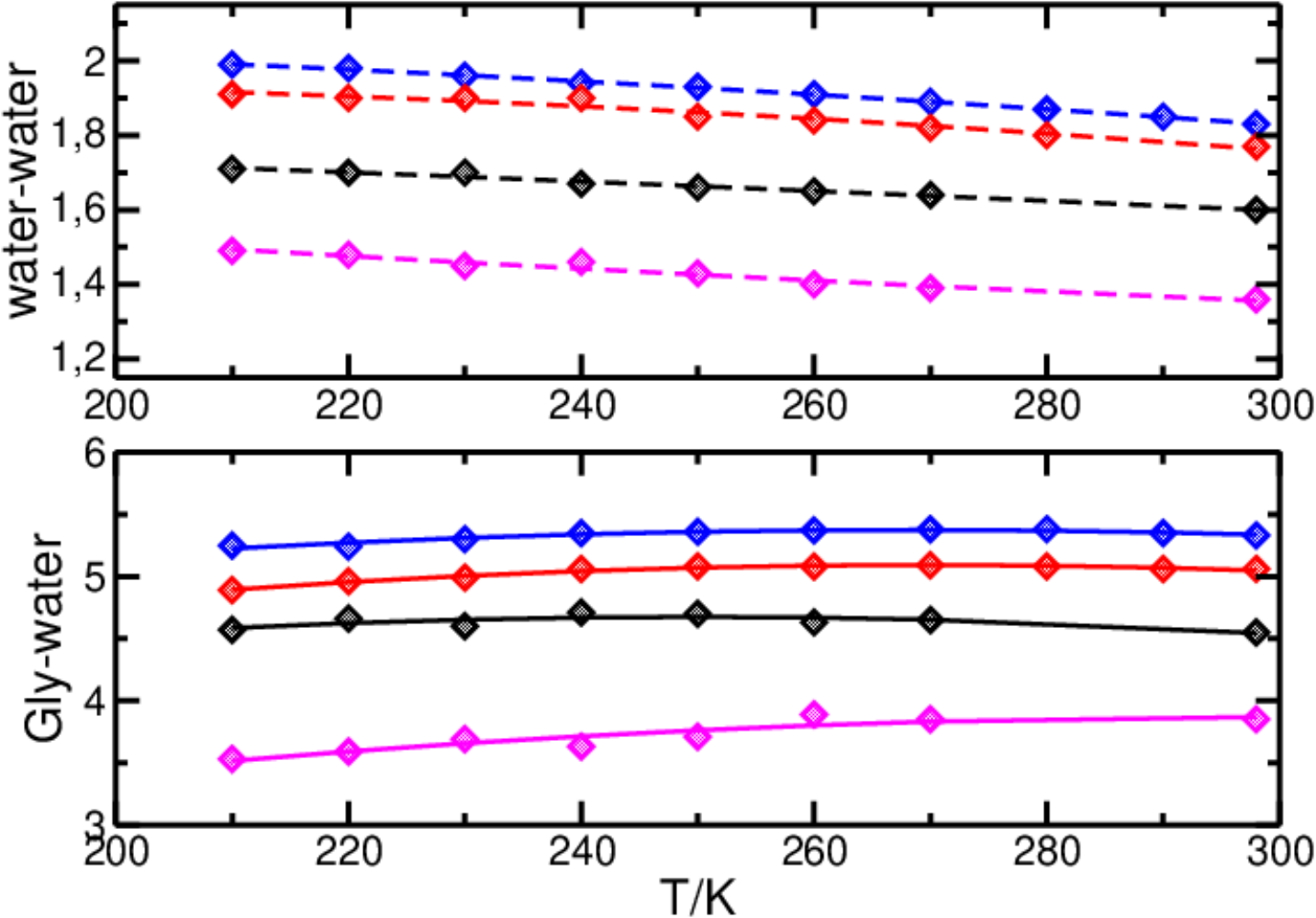}
    \caption{Average number of water-water hydrogen bonds formed per water molecule (top) and number of water-glycerol hydrogen bonds formed by each glycerol molecule (bottom).} 
    \label{fig:hbonds} 
\end{figure}

Finally, we focus our attention on the number of hydrogen bonds formed between water molecules and between water and glycerol.
In Figure \ref{fig:hbonds} we represent the average number of hydrogen bonds formed by a molecule at each concentration.

The top panel, representing the water-water hydrogen bonds,  agrees with previous calculations performed for pure water where the mean hydrogen bonds formed by one water molecule is roughly 2.\cite{guardia2006hydrogen} The number of hydrogen bonds decreases when increasing the glycerol concentration. 
\textcolor{black}{The bottom panel shows the number of hydrogen bonds that each glycerol molecule  forms with the surrounding water molecules. We divide the  total number of hydrogen bonds calculated by the number of glycerol molecules in the simulation box, therefore, the maximum number of hydrogen bonds that we may find per glycerol is $\approx$~5.5 in good agreement with previous results \cite{charkhesht2019insights}.
In the top panel, we observe a monotonic decrease in water-water bonds, however,  the bottom panel shows that the number of hydrogen bonds between glycerol and water remains constant at all temperatures. 
with the exception of $\chi_{g}$ = 50\% which seems to grow from 3.5 at T = 210 K to $\sim$ 4 at T = 298 K.  
As an hypothesis to explain this unexpected behaviour, one could think that at so high concentrations the solution is becoming a gel-like system where the glycerol rotation is restricted, which limits the number hydrogen bonds that can be formed. When increasing the temperature, the glycerol molecules gain flexibility increasing the probability of forming hydrogen bonds with the surrounding water molecules.}





  

\section{Conclusions}

In the present study we unravel the behaviour of transport properties and density anomaly of water-glycerol mixtures  at low temperatures using molecular dynamics simulations. We find an unexpected crossover in the dynamics of the 1\% and 10\% glycerol solutions that manifests across all properties tested: water's hydrogen bond lifetime, diffusion coefficients of water and glycerol, and viscosity of the solution. The dynamical crossover in all these properties occurs consistently around 220 K, and is observed in both the all-atom and coarse-grained models investigated in this work.

We find that the Stokes-Einstein relation for 1\% solution is strongly violated when considering either the diffusion coefficient of water or glycerol. Interestingly, the violation for water in the dilute solution is stronger than for pure water, despite the indistinguishable location of the anomalies in these two liquids. The implication is that the presence of an even small amount of glycerol increases the dynamical heterogeneities of the mixture (cluster formation), even though it does not have a noticeable effect on its thermodynamics. 

This is strongly dependent on temperature, and increases around the Widom line. The temperature evolution of the density of the 1\% glycerol solution suggests that its Widom temperature is close to the one of pure water. For the 10\% solution, on the other hand, the density maximum and Widom temperature moves down compared to pure water, 
shifting the conversion into four-coordinated water to lower temperatures.

We interpret that the dynamical crossover originates in the synergism of two factors that slow down the 1\% and 10\%  water-glycerol solutions. 
One contribution to  the slowing down is the glycerol-water and  gylcerol-glycerol interactions upon increasing the solution concentration. 
Another contribution to  the slowing down 
is the structural transformation of liquid water into an increasingly more four-coordinated liquid.

\textcolor{black}{
On the one side, the dynamical crossover might also be expected for other polyols. 
It is possible that, in polyols, the length of the molecule modifies the temperature at which this anomaly appears but more work would be needed in this direction. 
On the other side, we do not expect to detect this anomaly in certain ionic chaotropes such as NaCl. Although in both species (polyols and ions) we observe a displacement of the TMD\cite{corradini2011liquid}, the dynamics in alcohol mixtures  is strongly affected by the presence of the polyol, 
while for ions this effect is not so pronounced. Due to this effect, the observed crossover  in the dynamic properties at low temperatures should  not be present in salty solutions.}

\section*{Acknowledgement(s)}
This work is dedicated to Jose Luis Abascal, with whom most of the authors have collaborated in the last 12 years. Working with Jose Luis has been a pleasure and a  honour not only for being Jose Luis a recognised scientist, but also  a very modest and excellent person. His wise scientific and personal advises have been incredibly helpful for many of his collaborators and students who still learn from him nowadays.

\section*{Funding}
C.Valeriani acknowledges funding IHRC22/00002 and PID2022-140407NB-C21 from MINECO, R.K. and V.M. gratefully acknowledge
support by the U.S. Air Force Office of Scientific Research through a Multidisciplinary University Research Initiatives (MURI) Award FA9550-20-1-0351. FC acknowledges support from Agence Nationale de
la Recherche, Grant No. ANR-19-CE30-0035-01. The authors acknowledge the computer resources and technical assistance provided by the RES with funding FI-2023-2-0021 and the Center of High Performance Computing at the University of Utah. 
\bibliographystyle{tfo}
\bibliography{referencias}


\appendix

\section{}

\subsection{LAMMPS input files for the coarse grained model}
We provide the LAMMPS input files needed to simulate the coarse-grained glycerol-water model, whose details will be given in \cite{Katuaprabir2024}.


In table \ref{CGglycerolparam} we present the values for the non-bonded interactions between glycerol molecules calculated for the united-atom model. These interactions are Lennard-Jones type, while the O$_{gly}$-O$_{gly}$ interactions are Sillinger-Weber type and are presented in the main text.

\begin{table}[h!]
\tbl{Non-bonded interactions used in the nited atom model of glycerol. These interactions are Lennard-Jones type. These data will be published in ref. \cite{Katuaprabir2024}.}
{\begin{tabular}{c c c ccc | c ccccc c c} 
\hline
\multicolumn{4}{c}{Atom types} &&& &&&& \multicolumn{4}{c}{Parameters}   \\ 
\hline
\multicolumn{2}{c}{i} & \multicolumn{2}{c}{j} &&&&&&& \multicolumn{2}{c} {$\epsilon_{ij}$ (kcal $mol^{-1}$)} & \multicolumn{2}{c} { $\sigma_{ij}$ ($\AA$)} \\ 
\hline
\hline
 \multicolumn{2}{c} {CC} & \multicolumn{2}{c} {CC} &&&&&&& \multicolumn{2}{c} {0.080} & \multicolumn{2}{c} {4.043} \\
 \multicolumn{2}{c} {CC} & \multicolumn{2}{c} {CT} &&&&&&& \multicolumn{2}{c} {0.097} & \multicolumn{2}{c} {4.071} \\
 \multicolumn{2}{c} {CC} & \multicolumn{2}{c} {O} &&&&&&& \multicolumn{2}{c} {0.17} & \multicolumn{2}{c} {3.536} \\
 \multicolumn{2}{c} {CT} & \multicolumn{2}{c} {CT} &&&&&&& \multicolumn{2}{c} {0.118} & \multicolumn{2}{c} {4.100} \\
   \multicolumn{2}{c} {CT} & \multicolumn{2}{c} {O} &&&&&&& \multicolumn{2}{c} {0.17} & \multicolumn{2}{c} {3.536}\\
 \hline
\end{tabular}}
\label{CGglycerolparam}
\end{table}

\subsection{Glycerol diffusion and cluster formation}

Snapshot showing clustering for the coarse grained system at $\chi_g$ = 30 \% at T = 210K. We observe that at higher concentrations and low temperatures glycerol form clusters. These clusters decrease the diffusion of glycerol and water molecules.
\begin{figure}[h!]
\centering
    \includegraphics[width=0.3\columnwidth]{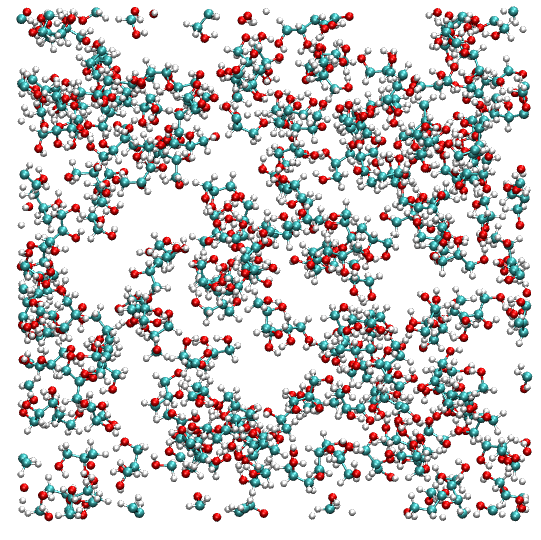}
    \caption{Snapshot of the system for $\chi_g$ = 30 \% at T = 210K. Glycerol at low temperatures and so high concentrations tends to form clusters.}
    \label{fig:clusters}
\end{figure}

\begin{table}[h!]
    \tbl{Glycerol diffusion  for 1\%,10\%, 30\% and 50\%  concentrations.}
    {\begin{tabular}{ccc|c|c|c|c|}
    \hline
     \multicolumn{5}{c}{D$_{glycerol}$ ($\mu m^{2}$/ s) } \\
    \hline 
T(K) &&& \multicolumn{1}{|c|}{1\%}& \multicolumn{1}{|c|}{10\%} & \multicolumn{1}{|c|}{30\%} &\multicolumn{1}{|c|}{50\%}    \\
    \hline
     \hline
298 &&& 791.836& 497.229 & 190.379 & 30.077 \\
270 &&&  352.987 & 200.114 & 66.8 & 4.995\\
260 &&&  217.386 & 124.6 & 33.648 & 2.249 \\
250 &&&  124.6 & 74.733 & 17.816 & 0.732\\ 
240 &&&  66.468 & 34.071&  7.351 &  0.295\\
230 &&&  21.836 & 13.542 & 2.959 & 0.122 \\
220 &&& 5.579 & 3.941 & 1 & 0.06 \\
210 &&& 0.832 &  0.819& 0.257 & 0.022\\
   \hline
     
     \end{tabular}}
       \label{tab:glycdiffap}
\end{table}






\subsection{Values of the diffusion for water and glycerol (all-atom)}

\begin{table}[h!]
\tbl{Water diffusion  for 1\%,10\% and 30\% glycerol concentrations. Pure water values  from \cite{montero2018viscosity}.*Interpolated values.}
{\begin{tabular}{cccccc}
    \hline
     \multicolumn{6}{c}{D$_{water}$ ($\mu m^{2}$/ s) } \\
    \hline 
 \multicolumn{1}{c}{T(K)} & \multicolumn{1}{c}{Water} & \multicolumn{1}{c}{1\%}& \multicolumn{1}{c}{10\%} & \multicolumn{1}{c}{30\%} &\multicolumn{1}{c}{50\%}    \\
    \hline
     \hline
298 & 2358.94* & 2296.415& 1697.303& 761.795& 186.782 \\
270 & 1039.4123* &1015.23& 706.342 & 268.463 & 41.826\\
260 & 701 &675.028& 462.602 & 173.186 & 23.197 \\
250 & 442.131* &403.763& 289.539 & 96.049 & 41.286\\ 
240 & 255 &210.791 & 146.657&  51.683 &  11.229\\
230 & 129.352*  &83.833 & 63.863 & 23.55 & 3.506 \\
220  & 57.6 &20.555 & 22.168& 8.554 & 1.55 \\
210 & - &2.925&  5.354& 2.514 & 0.538\\
   \hline
     \end{tabular}}
    \label{tab:watdiffa}
\end{table}





\end{document}


\section{Appendix}

\begin{figure}[h!]
    \includegraphics[width=1.0\columnwidth]{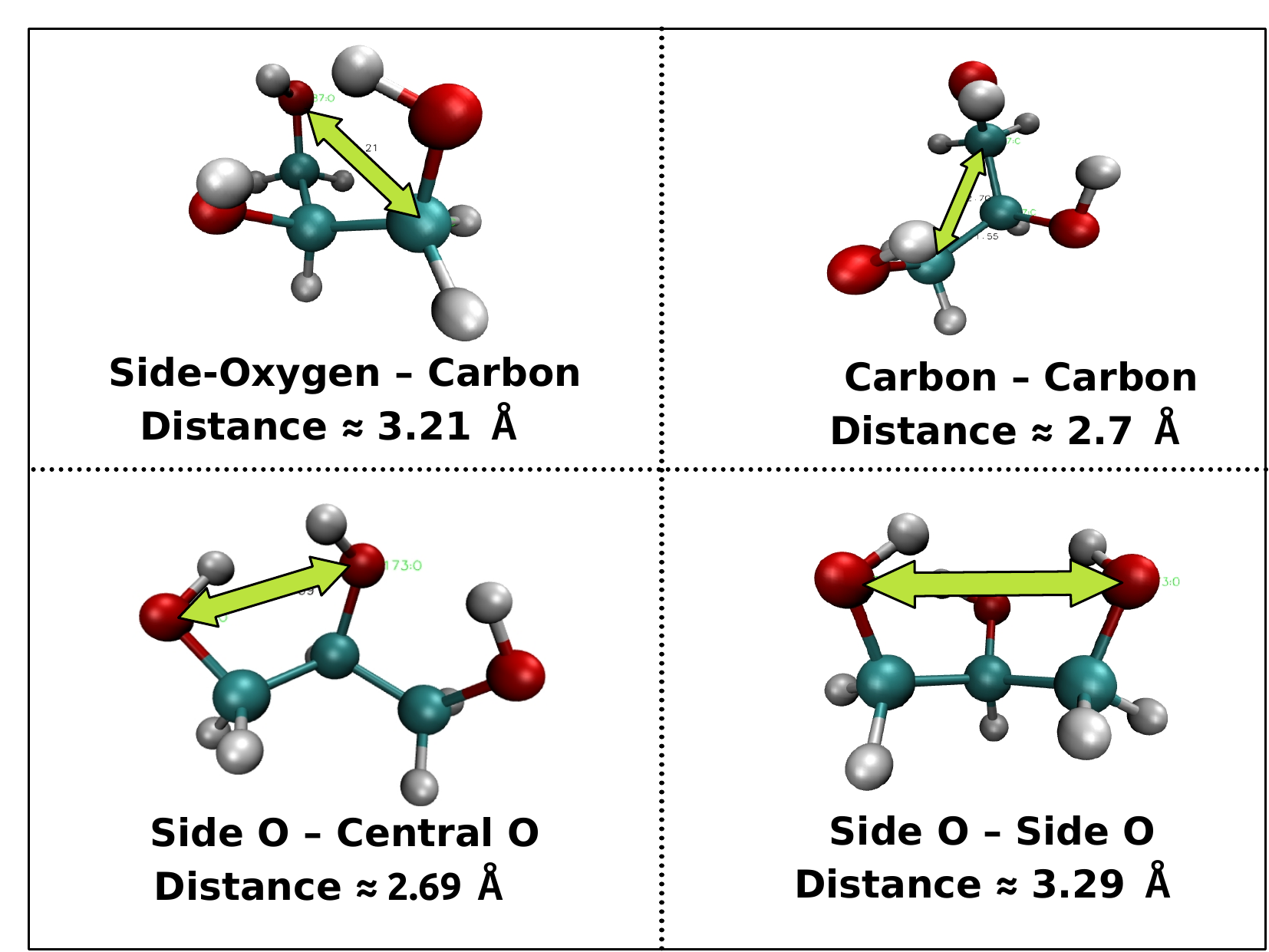}
    \caption{Intramolecular distances for glycerol molecule using BC-FF forcefield.}
    \label{fig:rdf_Ogly_OW1_all}
\end{figure}

\subsection{O$_{gly}$ $-$ O$_{water}$  Interaction}[h!]

To understand whether the system is always in a liquid state, we numerically compute the radial distribution function for BC-FF force-field $O_{water}$ - $O_{glycerol}$
 (fig.\ref{fig:rdf_Ogly_OW1_all}) when the concentration is 1\% (top) and 10\% (bottom) for a mixture at 220K (red) and 298K (blue)

\begin{figure}[h!]
    \includegraphics[width=1.0\columnwidth]{rdf_Ogly_OW1_all.eps}
    \caption{Radial distribution function for oxygen between O$_{glycerol}$-O$_{water}$ computed from T = 210 to 298 K indicated for concentrations Xg = 1 \% (blue), Xg = 10 \% (red), Xg = 30 \% (black) and Xg = 50 \% (magenta).}
    \label{fig:rdf_Ogly_OW1_all}
\end{figure}

\begin{figure}[h!]
    \centering
    \includegraphics[width=1.0\columnwidth]{max_min_rdf_OW-Ogly.eps}
    \caption{Top: Position of first, second and third maximum peaks found for RDF as a function of temperature. Bottom: Position of the first, second and third minimun of RDF versus temperature calculated for the pair O_{water}-O_{glycerol}.}
\label{fig:maxminglywat} 
\end{figure}

\begin{figure}[h!]
    \centering
        \includegraphics[width=1.0\columnwidth]{integral_RDF_Ogly_OW1_0_1st_min.eps}
    \caption{Number of neighbours calculated from 0 to the 1st minimum of the RDF for O$_{glycerol}$-O$_{water}$ at all the temperatures indicated for concentrations Xg = 1 \% (black),Xg = 10\% (red), Xg = 30\% and 50 \% . }
    \label{fig:1neighbglywat} 
\end{figure}

\begin{figure}[h!]
    \centering
        \includegraphics[width=1.0\columnwidth] {integral_RDF_Ogly_OW1_1st_min_to_2nd_min.eps}
    \caption{Number of neighbours calculated from 1st minimum to the 2nd minimum of the RDF for O$_{glycerol}$-O$_{water}$ at all the temperatures indicated for concentrations Xg = 1 \% (black),Xg = 10\% (red), Xg = 30\% and 50 \% . }
    \label{fig:2neighbglywat} 
\end{figure}

\begin{figure}[h!]
    \centering
        \includegraphics[width=1.0\columnwidth]{integral_RDF_Ogly_OW1_2ndmin_3rd_min.eps}
    \caption{Number of neighbours calculated from  the 2nd minimum to the 3rd min of the RDF for O$_{glycerol}$-O$_{water}$ at all the temperatures indicated for concentrations Xg = 1 \% (black),Xg = 10\% (red), Xg = 30\% and 50 \% . }
    \label{fig:3neighbglywat} 
\end{figure}


\subsection{O$_{gly}$ $-$ O$_{gly}$  Interaction}

\begin{figure}[h!]
    \includegraphics[width=1.0\columnwidth]{RDF-intermolecular_Ogly_Ogly_all.eps}
    \caption{Radial distribution function for oxygen between O$_{glycerol}$-O$_{glycerol}$ computed from T = 210 to 298 K indicated for concentrations Xg = 1 \% (blue), Xg = 10 \% (red), Xg = 30 \% (black) and Xg = 50 \% (magenta).}
    \label{fig:rdf_Ogly_Ogly_all}
\end{figure}

\begin{figure}[h!]
    \centering
    \includegraphics[width=1.0\columnwidth]{max_min_Ogly_Ogly.eps}
    \caption{Top: Position of first, second and third maximum peaks found for RDF as a function of temperature. Bottom: Position of the first, second and third minimun of RDF versus temperature calculated for the pair O_{glycerol}-O_{glycerol}.}
\label{fig:maxminglygly} 
\end{figure}

\subsection{O$_{gly}$ $-$ C  Interaction}

\begin{figure}[h!]
    \includegraphics[width=1.0\columnwidth]{RDF-Ogly-C.eps}
    \caption{Radial distribution function for oxygen between O$_{glycerol}$-Carbon computed from T = 210 to 298 K indicated for concentrations Xg = 1 \% (blue), Xg = 10 \% (red), Xg = 30 \% (black) and Xg = 50 \% (magenta).}
    \label{fig:rdf_Ogly_C_all}
\end{figure}

\begin{figure}[h!]
    \includegraphics[width=1.0\columnwidth]{max-min-Ogly-C.eps}
    \caption{Top: Position of first, second and third maximum peaks found for RDF as a function of temperature. Bottom: Position of the first, second and third minimun of RDF versus temperature calculated for the pair O_{glycerol}-Carbon.}
    \label{fig:maxmin_Ogly_C_all}
\end{figure}

\subsection{Carbon $-$ Carbon  Interaction}

\begin{figure}[h!]
    \includegraphics[width=1.0\columnwidth]{RDF-C-C.eps}
    \caption{Radial distribution function between Carbon-Carbon computed from T = 210 to 298 K indicated for concentrations Xg = 1 \% (blue), Xg = 10 \% (red), Xg = 30 \% (black) and Xg = 50 \% (magenta).}
    \label{fig:rdf_C_C_all}
\end{figure}

Here we will compare RDF for BC-FF and 2FF.




\begin{table}[h!]
    \centering
    \begin{tabular}{|c|c|c|c|c|c|}
    \hline
     \multicolumn{6}{|c|}{D_{water} (\mu m^{2}/ s) } \\
    \hline 
 \multicolumn{1}{|c|}{T(K)} & \multicolumn{1}{|c|}{Water} & \multicolumn{1}{|c|}{1\%}& \multicolumn{1}{|c|}{10\%} & \multicolumn{1}{|c|}{30\%} &\multicolumn{1}{|c|}{50\%}    \\
    \hline
     \hline
298 & 2358.94* & 2296.415& 1697.303& 761.795& 186.782 \\
270 & 1039.4123* &1015.23& 706.342 & 268.463 & 41.826\\
260 & 701 &675.028& 462.602 & 173.186 & 23.197 \\
250 & 442.131* &403.763& 289.539 & 96.049 & 41.286\\ 
240 & 255 &210.791 & 146.657&  51.683 &  11.229\\
230 & 129.352*  &83.833 & 63.863 & 23.55 & 3.506 \\
220  & 57.6 &20.555 & 22.168& 8.554 & 1.55 \\
210 & - &2.925&  5.354& 2.514 & 0.538\\
   \hline
     \end{tabular}
    \caption{Water diffusion computed for 1\%,10\% and 30\% glycerol concentrations. The water diffusion values were obtained from \cite{montero2018viscosity}.*Interpolated values.}
    \label{tab:watdiffa}
\end{table}

\begin{table}[h!]
    \centering
    \begin{tabular}{|c|c|c|c|c|}
    \hline
     \multicolumn{5}{|c|}{D_{glycerol} (\mu m^{2}/ s) } \\
    \hline 
 \multicolumn{1}{|c|}{T(K)} & \multicolumn{1}{|c|}{1\%}& \multicolumn{1}{|c|}{10\%} & \multicolumn{1}{|c|}{30\%} &\multicolumn{1}{|c|}{50\%}    \\
    \hline
     \hline
298 & 791.836& 497.229 & 190.379 & 30.077 \\
270 &  352.987 & 200.114 & 66.8 & 4.995\\
260 &  217.386 & 124.6 & 33.648 & 2.249 \\
250 &  124.6 & 74.733 & 17.816 & 0.732\\ 
240 &  66.468 & 34.071&  7.351 &  0.295\\
230 &  21.836 & 13.542 & 2.959 & 0.122 \\
220 & 5.579 & 3.941 & 1 & 0.06 \\
210 & 0.832 &  0.819& 0.257 & 0.022\\
   \hline
     \end{tabular}
    \caption{Glycerol diffusion computed for 1\%,10\% and 30\% glycerol concentrations.}
    \label{tab:glycdiffap}
\end{table}

\begin{table}[h!]
    \centering
    \begin{tabular}{|c|c|c|c|c|}
    \hline
     \multicolumn{5}{|c|}{Viscosity \eta (cp) - GK} \\
    \hline 
 \multicolumn{1}{|c|}{T(K)} & \multicolumn{1}{|c|}{Water} & \multicolumn{1}{|c|}{1\%}& \multicolumn{1}{|c|}{10\%} & \multicolumn{1}{|c|}{30\%}     \\
    \hline
     \hline
298 &0.87*& 0.86 & 1.24& 4.11  \\
270 & 2.05* & 2.05 & 2.83 & 10.48 \\
260 & 2.753 &2.69 & 4.46 & 18.31  \\
250 & 4.153* &5.26 & 6.84 & 32.86  \\ 
240 & 7.94 &8.95 & 14.24 & 62.86\\
230 & 17.52*  &26.78 & 33.94 & 156.49 \\
220  & 36.8 &137.31 & 130.76& 510.68 \\
210 & - &2136.16&  598.08 & 1725.6 \\
   \hline
     \end{tabular}
    \caption{Viscosity computed by means of Green-Kubo relation for 1\%, 10\% and 30\% concentrations. The values for pure water were obtained from \cite{montero2018viscosity } *Interpolated values. }
    \label{tab:my_label}
\end{table}

\begin{table}[h!]
    \centering
    \begin{tabular}{|c|c|c|c|c|c|}
    \hline
     \multicolumn{6}{|c|}{Viscosity \eta (cp) - SE} \\
    \hline 
 \multicolumn{1}{|c|}{T(K)} & \multicolumn{1}{|c|}{Water} & \multicolumn{1}{|c|}{1\%}& \multicolumn{1}{|c|}{10\%} & \multicolumn{1}{|c|}{30\%} & \multicolumn{1}{|c|}{50\%} \\
    \hline
     \hline
298 &0.845& 0.868 & 1.174& 2.617 & 10.672  \\
270 & 1.875 & 1.919  & 2.758 & 7.258  & 46.584 \\
260 & 2.786 & 2.894  & 4.222 &11.278& 84.202  \\
250 & 4.43 & 4.85 & 6.764 & 20.39  & 174.408 \\ 
240 & 7.746 & 9.37 & 13.468 & 38.217 & 357.954 \\
230 & 15.505 & 23.923   & 31.404 & 85.161 & 572.031 \\
220  & 35.189 & 98.609 & 91.434 & 236.954 & 1307.682  \\
210 & -        &715.474&  390.878 & 832.443 & 3889.891 \\
   \hline
     \end{tabular}
    \caption{Viscosity computed by means of Stokes-Einstein relation for 1\%, 10\%, 30\% and 50\% concentrations. The values for pure water were obtained from \cite{montero2018viscosity } *Interpolated values. }
    \label{tab:my_label}
\end{table}